ARTICLE

# Adtech and Real-Time Bidding under European Data Protection Law

Michael Veale[1],  and Frederik Zuiderveen Borgesius[2]

[1]Associate Professor, Faculty of Laws, University College of London, London, United Kingdom and [2]Professor, Radboud University, Nijmegen, The Netherlands
Corresponding Author Email: m.veale@ucl.ac.uk

## Abstract

This article discusses the troubled relationship between contemporary advertising technology (adtech) systems, in particular systems of real-time bidding (RTB, also known as programmatic advertising) underpinning much behavioral targeting on the web and through mobile applications. This article analyzes the extent to which practices of RTB are compatible with the requirements regarding a legal basis for processing, transparency, and security in European data protection law.

We first introduce the technologies at play through explaining and analyzing the systems deployed online today. Following that, we turn to the law. Rather than analyze RTB against every provision of the General Data Protection Regulation (GDPR), we consider RTB in the context of the GDPR's requirement of a legal basis for processing and the GDPR's transparency and security requirements. We show, first, that the GDPR requires prior consent of the internet user for RTB, as other legal bases are not appropriate. Second, we show that it is difficult—and perhaps impossible—for website publishers and RTB companies to meet the GDPR's transparency requirements. Third, RTB incentivizes insecure data processing. We conclude that, in concept and in practice, RTB is structurally difficult to reconcile with European data protection law. Therefore, intervention by regulators is necessary.

## A. Introduction

This article discusses the troubled relationship between EU data protection legislation, encompassing the EU General Data Protection Regulation (GDPR) and ePrivacy instruments, and the infrastructures of contemporary behavioral targeting. Behavioral targeting is the monitoring of online behavior, and the use of this to deliver personalized advertisements. Today, both on the Web and in packaged software, such as mobile apps, a complex, interwoven web of actors and technologies operate in concert to deliver the granular, and often uncanny, tailoring seen today. A main mode of online advertising today is known as *real-time bidding* (RTB). This article focuses quite simply on the following question: To what extent is real-time bidding compatible with EU data protection and privacy law's requirements regarding a legal basis for processing, transparency, and security?

Addressing this question requires a synthesis of empirical work in computer science with a careful consideration of the current data protection regime and varying authorities. The answer is consequential, given that a majority of European citizens will have interacted with various RTB systems in recent years, if not in recent weeks or days alone. Section 2 introduces advertising technologies in a historical and technological context. It outlines the technologies and practices underpinning the RTB system. We then turn to the law. We argue that the GDPR generally applies to RTB. We show that the GDPR requires consent of the internet user as a legal basis for real-time bidding practices, while the ePrivacy Directive also requires consent. Next, we show that it would be extremely difficult to make RTB comply with the GDPR's transparency requirements and security requirements. We briefly discuss the findings before we conclude by calling upon regulators to enforce the GDPR in the RTB sector.

## B. Online Advertising, Adtech, and RTB

RTB is a system where pre-determined advertising space, such as a banner advert on a website, or a splash screen in an app, is allocated through an auction process carried out for each requested impression. The creation of markets does not directly engage privacy concerns. Advertising can be envisaged without the use of personal data. Potential advertising space can be auctioned on the basis of generic data which does not individuate a viewer, such as the time of day, the country of internet access, the content of the page the advert is shown on, and so on.[1]

In practice however, those participating in auctions for online advertising do not only consider characteristics of the property—for example, the website's content—but evaluate the personal data of the user. RTB is heavily entwined with individualized tracking and cannot be properly understood without it.[2] We therefore explain the underlying infrastructure, before elaborating further on the functioning of the RTB system.

### I. Online Tracking

Tracking infrastructure can be split into two main types, *explicit* tracking and *inferred* tracking.[3] These terms refer to the role of the user, their device, and unique identifiers, in the tracking process.

#### 1. Explicit Tracking

Explicit tracking occurs when a user is identified by a tracking mechanism that assigns a unique identifier per user. A user's device may emit a unique identifier for functional reasons which can be used to track them, or a tracking infrastructure may have caused the device to store and emit an identifier on request.

On the Web, cookies are one of the main methods used in explicit tracking. Cookies were invented in 1994 to provide an ability to retain state to the stateless HTTP protocol, and "[give] the Web a memory."[4] They consist of text, often encoded or even encrypted,[5] that can be placed by a server in a user's browser and examined later by a server. Cookies have a range of useful functions; without some ability for a website to store information in a browser, login status, for example, could not be reliably remembered the next time a user visits a site. It would also be challenging for e-commerce applications, such as retaining items in a basket online. However, the seamless and silent nature of cookies has also meant cookies can be used in ways that go beyond users' expectations.

---

[1]Such advertising is often called "contextual advertising."

[2]The real-time bidding industry is uncomfortable with the term tracking—calling its condemnation "facile"—but does not deny that it describes the practices they undertake. *See* Helen Mussard, *Digital Advertising Industry Warns Against Misguided EU Regulations*, INTERACTIVE ADVERTISING BUREAU EUROPE, Sept. 29, 2020, https://perma.cc/GQ6J-GXVW.

[3]Franzisk Roesner, Tadayoshi Kohno, and David Wetherall, *Detecting and Defending Against Third-Party Tracking on the Web*, 9TH USENIX SYMP. ON NETWORKED SYS. DESIGN & IMPLEMENTATION, 158 (2012),
https://www.usenix.org/system/files/conference/nsdi12/nsdi12-final17.pdf.

[4]John Schwartz, *Giving the Web a Memory Cost Its Users Privacy*, N.Y.TIMES, Sept. 4, 2001, https://www.nytimes.com/2001/09/04/business/giving-web-a-memory-cost-its-users-privacy.html.

[5]For example, IAB cookies relating to real-time bidding are often placed as concatenated, sequential bits that are then encoded using the base64 method.

In the early days of the Internet—when nobody knew which users were dogs[6]—content on webpages usually only came from a single source. Netscape Navigator 2.0 introduced the function of rendering two HTML files in a single browsing window in 1996 through *frames*, and so security features were needed to determine which frame could access which information in the browser. The *Same Origin Policy*, broadly put, means that documents in the browser, such as cookies, can only be accessed by servers sharing their protocol, for example, HTTP or HTTPS, domain, and port.[7] The intention for this was so that if one party places a cookie, another party cannot read it.

Driven by a desire to establish cross-site tracking, the advertising industry sought to circumvent the effects of the Same Origin Policy. Cookies that are not placed by the website publisher itself are often called *third-party cookies*. Say that somebody visits a website, www.A.com. It may seem that each element on that website comes from A.com. In reality, however, different elements of the website are often sourced from other domains. For example, a website may have a box for advertisements, or for recommended articles elsewhere on the web. In most cases, ads are shown on a website not by the website publisher itself—A, in our example—but by third parties. Those third parties can also place and read their own cookies, third party cookies. The website visitor usually does not see that their browser contacts these different domains; for the website visitor it seems like one website is being loaded from one domain. While a user may only see one URL in the address bar, visiting almost any site now entails querying tens or hundreds of other servers.

Some firms have spread their own tracking code with resounding success: Google calls home with unique identifiers for at least twenty-eight percent of all web page loads, while Facebook does the same for approximately fifteen percent.[8] The proportion is significantly higher in certain sectors, such as news, compared to others, such as banking. Trackers also differ by country—U.K. users are tracked more in this manner than Chinese web users, for example.[9]

Yet firms with less infrastructure also established means to track users more broadly by using loopholes in the Same Origin Policy to combine the reach of their tracking. The prime mechanism this is carried out is through cookie syncing, also called cookie matching. In its most basic form, this involves a third-party with a cookie ("TRACKER1") making a user's browser query a second third-party ("TRACKER2") with a URL which includes TRACKER1's identifier.[10] Because the user's browser is querying TRACKER2, TRACKER2 is able to look at its own cookies on the site. As the query includes the ID that TRACKER1 just saw from its own cookies, this has the effect of enabling TRACKER2 to possess both identifiers at once, associating their own cookie ID with TRACKER1's cookie ID. The two organizations can share data through a back-channel server-to-server transfer[11] to connect the profiles they have built so far. This cookie syncing significantly widens the scope of tracked activity online by pooling the reach of multiple trackers.[12] Even under conservative estimates of server-to-server transfers—based only on observed cookie syncing—fifty three firms observe more than ninety-one percent of users' browsing behavior.[13] This figure is likely an underestimation; a recent study found evidence that as many as twenty-seven percent of advertiser-tracker relationships may be undetectable through cookie syncing analysis.[14]

---

[6]*See generally* Glenn Fleishman, *Cartoon Captures Spirit of the Internet*, N.Y. TIMES, Dec. 14, 2000, https://www.nytimes.com/2000/12/14/technology/cartoon-captures-spirit-of-the-internet.html.

[7]*See generally* Frederik Braun, *Origin Policy Enforcement in Modern Browsers*, Oct. 26, 2012, https://frederik-braun.com/publications/thesis/Thesis-Origin_Policy_Enforcement_in_Modern_Browsers.pdf.

[8]Arjaldo Karaj, Sam Macbeth, Remi Berson, and Josep M. Pujol, *WhoTracks. Me: Monitoring the Online Tracking Landscape at Scale*, (Apr. 2019), https://arxiv.org/abs/1804.08959.

[9]Xuehui Hu, Guillermo Suarez de Tangil, and Nishanth Sastry, *Multi-Country Study of Third Party Trackers from Real Browser Histories*, 2020 IEEE EUR. SYMP. ON SEC. & PRIVACY 70 (2020).

[10]For example, a query might look like http://tracker2.com?tracker1cookieID=j9240.

[11]*See generally* Muhammad Ahmad Bashir and Christo Wilson, *Diffusion of User Tracking Data in the Online Advertising Ecosystem*, 2018 PROCEEDINGS ON PRIV. ENHANCING TECH., 85 (Jan. 2018).

[12]Google calls this "cookie matching." *See generally* Jane Wakefield, *Google's 'secret Web Tracking Pages' Explained*, BBC, May 9, 2019, https://www.bbc.com/news/technology-49593830.

[13]Bashir and Wilson, *supra* note 11.

More recently, trackers have sought to evade restrictions on third-party cookies in a number of newer ways. In particular, they have been encouraging sites to edit their Domain Name System (DNS) records to effectively deliver third-party tracker resources from the same domain that is serving the website, making effectively blocking trackers without breaking the website a trickier task. This also creates a range of serious security risks, as first-party cookies often include cookies designed to log a user in to a website, and such configurations can mean that these cookies can be read by and sent to a third party other than the website operator.[15] As a result, the terms "first-party cookies" and "third-party cookies" have less meaning in a legal context; a case-by-case analysis will be required to understand which actors are involved in any particular cookie, as the domain name-based identity of the server laying it may not be the same as the organization utilizing its tracking potential across websites.

On mobile devices, app developers, which are analogous to website publishers, have more freedom to execute arbitrary code. As the Web is accessed through a browser, the browser has power to limit the ways a website can function. In contrast, app developers can specify the way their software works without being required to cede rendering and execution decisions to the browser. Instead, looser limits are applied at the level of the mobile operating system—for example, limiting access to sensors such as the camera or GPS—and through any conditions placed upon apps allowed to be distributed through official channels such as Apple's App Store and Alphabet's Play Store, which for most users will be the only way they install custom software.

The fact that app developers have more freedom than Web developers to determine who is able to track individuals means that active tracking, rather than passive tracking, is the main issue in the mobile sphere. Third-party services are integrated in apps for a variety of purposes, including crash reporting, to provide usage or engagement analytics, to integrate agile development methods such as A/B testing, to integrate with other services such as social networks, and to deliver advertising. Almost all of these services, with the exception of advertising services, only operate in the background of applications, and users are in general unable to detect and understand the extent to which the app is communicating with both the developer's server or third-party servers.

Empirical studies into tracking apps are challenging and are mostly limited to the Android platform due to the inability to examine the innards of the heavily restrictive Apple iOS system. Studies seeking to survey app tracking at scale take a few different approaches.[16] Some researchers intercept traffic from hundreds of thousands of apps which are being either interacted with automatically by bots synthesizing real user input in a sandbox on a server,[17] or by using real user interactions, with traffic captured via user-installed VPNs.[18] One recent study identified 2,121 separate advertising tracking services in apps in the Android ecosystem, which can be grouped by ownership into approximately 292 parent organizations.[19] Another study found that 88.4 percent of apps contained a tracker owned by Alphabet (Google), 42.6 percent by Facebook, 33.9 percent by Twitter, 26.3 percent by Verizon and 22.2 percent by Microsoft.[20] Thirty-percent of News apps, twenty-eight percent of Family apps, and twenty-five percent of Gaming & Entertainment apps contain trackers from more than ten distinct tracker companies.[21]

---

[14]John Cook, Rishab Nithyanand, and Zubari Shafiq, *Inferring Tracker-Advertiser Relationships in the Online Advertising Ecosystem Using Header Bidding*, 2020 PROCEEDINGS ON PRIV. ENHANCING TECH., 65 (Jan. 2020).

[15]Yana Dimova, Gunes Acar, Lukasz Olejnik, Wouter Joosen, and Tom Van Goethem, *The CNAME of the Game: Large-Scale Analysis of DNS-Based Tracking Evasion*, Mar. 5, 2021, https://arxiv.org/abs/2102.09301.

[16]For an assessment of the comparative merits of these approaches, see Abbas Razaghpanah, Rishab Nithyanand, Narseo Vallina-Rodriguez, Srikanth Sundaresan, Mark Allman, Christian Kreibich, and Phillipa Gill, *Apps, Trackers, Privacy, and Regulators: A Global Study of the Mobile Tracking Ecosystem*, THE NETWORK & DISTRIBUTED SYS. SEC. SYMP., 13–14 (Feb. 2018).

[17]See, e.g., Haojian Jin, Minyi Liu, Kevan Dodhia, Yuanchun Li, Gaurav Srivastava, Matthew Fredrikson, Yuvraj Agarwal, and Jason I. Hong, *Why Are They Collecting My Data?: Inferring the Purposes of Network Traffic in Mobile Apps*, 2 PROC. ACM INTERACT. MOBILE WEARABLE UBIQUITOUS TECH. 1, 3–4 (Dec. 2018).

[18]See, e.g., Razaghpanah, supra note 16, at 3; Anastasia Shuba, Anh Le, Emmanouil Alimpertis, Minas Gjoka, and Athina Markopoulou, *AntMonitor: A System for On-Device Mobile Network Monitoring and Its Applications*, (Apr. 2017), https://arxiv.org/abs/1611.04268; Yihang Song and Urs Hengartner, *PrivacyGuard: A VPN-Based Platform to Detect Information Leakage on Android Devices*, PROCS. OF THE 5TH ANN. ACM CCS WORKSHOP ON SEC. & PRIV. IN SMARTPHONES AND MOBILE DEVICES, 15 (2015).

Mobile devices hold a variety of unique identifiers tied to their software and hardware with different levels of permanence, such as the IMEI, IMSI and SIM number, operating system number, phone number, device ID, MAC address, and operating system-specific advertising identifiers.[22] Third-party plug-ins have a variety of direct and indirect ways to access these identifiers, and in practice access and transmit a wide variety of them.[23] Such identifiers are also linked through a variety of means to track individuals across different devices, although exactly how this occurs in-the-wild is unclear.[24]

### 2. Inferred Tracking

Inferred tracking seeks to identify or profile an individual from observing their digital traces online and re-identifying a user through primarily probabilistic means. Unlike explicit tracking, these approaches are "stateless"—they do not change the behavior of user's devices, nor store information on them directly. Inferred tracking is therefore substantially more challenging for an individual or device to defend against.

Fingerprinting is a core approach for inferred tracking. Early documentation of fingerprinting was provided by analysis from the Electronic Frontier Foundation's *Panopticlick* tool, which uses modern fingerprinting techniques to determine how unique—and therefore how fingerprintable—your browser is.[25] Browser fingerprinting is a moving target, as sophisticated techniques can circumvent proxies, reveal the particular version of a browser, and repurpose new Web technologies for fingerprinting as they emerge.[26] Research has found evidence of fingerprinting on at least 4.4 to 5.5 percent of top websites, although these should be taken as lower bounds due to the difficult-to-observe nature of fingerprinting techniques.[27] These methods interplay with explicit tracking mechanisms—if the user clears their cookies, for example, fingerprinting approaches can be used to re-establish or "respawn" deleted identifiers.[28]

---

[19]Razaghpanah, *supra* note 16, at 7.

[20]Reuben Binns, Ulrik Lyngs, Max Van Kleek, Jun Zhao, Timothy Libert, and Nigel Shadbolt, *Third Party Tracking in the Mobile Ecosystem*, PROCS. 10TH ACM CONF. ON WEB SCI. 23, 27 (2018). The sixth most prevalent tracker was LinkedIn, which has since been purchased by the fifth most prevalent tracker, Microsoft.

[21]Binns, *supra* note 20, at 28.

[22]Razaghpanah, *supra* note 16, at 3.

[23]*Id.* at 7.

[24]*See generally* Sebastian Zimmeck, Jie S. Li, Hyungtae Kim, Steven M. Bellovin, and Tony Jebara, *A Privacy Analysis of Cross-Device Tracking*, 26TH USENIX SEC. SYMP. (2017), https://www.usenix.org/conference/usenixsecurity17/technical-sessions/presentation/zimmeck.

[25]*See* Peter Eckersley, *How Unique Is Your Web Browser?*, *in* PRIVACY ENHANCING TECHNOLOGIES 14 (Mikhail J. Atallah and Nicholas J. Hopper eds., 2010). You can test your own browser at https://panopticlick.eff.org.

[26]*See generally* Nick Nikiforakis, Alexandros Kapravelos, Wouter Joosen, Christopher Kruegel, Frank Piessens, and Giovanni Vigma, *Cookieless Monster: Exploring the Ecosystem of Web-Based Device Fingerprinting*, 2013 IEEE SYMP. ON SEC. & PRIVACY 541 (May 2013); Łukasz Olejnik, Gunes Aca, Claude Castelluccia, and Claudia Diaz, *The Leaking Battery*, 9481 DATA PRIV. MGMT. SEC. ASSURANCE 254 (2015).

[27]Gunes Acar, Marc Juarez, Nick Nikiforakis, Claudia Diaz, Seda Güres, Frank Piessens, and Bart Preneel, *FPDetective: Dusting the Web for Fingerprinters*, PROCEEDINGS OF THE 2013 ACM SIGSAC CONF. ON COMPUT. & COMMUNICATIONS SEC. 1129 (2013); Gunes Acar, Christian Eubank, Steven Englehardt, Marc Juarez, Arvind Narayanan, and Claudia Diaz, *The Web Never Forgets: Persistent Tracking Mechanisms in the Wild*, PROCS. OF THE 2014 ACM SIGSAC CONF. ON COMPUT. & COMMC'NS SEC. 674 (2014).

[28]Ashkan Soltani, Shannon Canty, Quentin Mayo, Lauren Thomas, and Chris Jay Hoofnagle, *Flash Cookies and Privacy*, 2010 AAAI SPRING SYMP. SERIES (Mar. 2010); Mika D. Ayenson, Dietrich James Wambach, Ashkan Soltani, Nathan Good,

Inferred tracking also plays an important role in cross-device tracking. Simulated cross-device tracking studies have estimated a significant ability to follow users from their mobiles to their desktops, even if they are not logged in to the same service—for example, if they are connected to the same router. Many companies advertise probabilistic cross-device tracking as a reason to install and use their trackers.[29] Despite scholarly interest, the covert, stateless nature of fingerprinting and inferred tracking makes its prevalence, scope, and affect unclear.[30]

## *II. RTB and Programmatic Advertising*

RTB is a form of programmatic advertising, where advertising placements are determined by algorithmic systems, rather than in human-mediated ways, such as through traditional negotiation and contracts. With RTB, advertisers—or their intermediaries—bid on an automated auction for the chance to target an ad to a specific internet user. RTB is also called audience selling or audience buying.

While early display advertising—the sale of "properties" such as banners, pop-ups or video segments—was largely conducted through manual deals, advertising is now predominately allocated automatically through programmatic methods, of which RTB is the prime system.[31]

RTB is a complicated system, with many different types of players. Below, we give a brief introduction to RTB. The reader might be a tad overwhelmed, but such a reaction is understandable. We return to the complexity and the opaqueness of RTB in section E.

In brief, RTB works as follows—and is illustrated in the companion Figure 1: A website or app publisher has a range of slots that it wishes to sell to advertisers, known in the industry as their "inventory." Advertisers to fill these slots are sought through one or more supply-side platforms (SSPs) the publisher deals with.[32] SSPs are technical intermediaries for publishers to work with complex advertising auction markets called advertising exchanges (ADXS), which serve as auction-houses for RTB. Demand-side platforms (DSPs) are the technical intermediaries that represent advertisers. Such DSPs place bids on ADXS.

These DSPs will be given a copy of the bid request, which represents information about the user to whom the advertisement will be delivered. This information varies slightly depending on the specification. On the Web, the contents of this bid request is determined by one of two specifications. The first is *Authorized Buyers*, a specification determined by Google. The second is *OpenRTB/AdCOM*, maintained by the technology division of the Interactive Advertising Bureau (IAB), a membership organization of large and small advertising firms ranging from Google, Facebook, and Twitter down to smaller actors.

---

Chris Jay Hoofnagle, *Flash Cookies and Privacy II: Now with HTML5 and ETag Respawning*, SSRN ELEC. J. (Jul. 2011), https://papers.ssrn.com/sol3/papers.cfm?abstract_id=1898390; Acar, *The Web Never Forgets*, supra note 27, at 675.

[29]Zimmeck et al., *supra* note 24.

[30]A useful review of knowledge on fingerprinting is Pierre Laperdrix, Nataliia Bielova, Benoit Baudry, and Gildas Avoine, *Browser Fingerprinting: A Survey*, (Nov. 2019), https://arxiv.org/abs/1905.01051.

[31]COMPETITIONS & MARKETS AUTH., ONLINE PLATFORMS AND DIGITAL ADVERTISING: MARKET STUDY FINAL REPORT ¶ 2.41 (July 2020), https://www.gov.uk/cma-cases/online-platforms-and-digital-advertising-market-study#final-report.

[32]In the past, a publisher would work with a single SSP. However, as are several AdXs, publishers often found themselves locked into a "waterfall" process whereby, in practice, Google's exchange would get the first attempt to bid for inventory, and if it met the minimum specifications (e.g., a price threshold) of publishers, it would be the ad that was served. Only if this failed would the same ad be offered to other ad exchanges, in a waterfall–like sequence. In some cases, these other exchanges may have offered a higher price, but the publisher would never have known. Consequently, in recent years, a process called *header bidding* emerged, where the publisher themselves (in the HTML header of the website) would query many exchanges simultaneously rather than sequentially, through multiple SSPs, and on their end (or the server of yet another intermediary) evaluate the different offers and choose between them. *See generally* Michalis Pachilakis, Panagiotis Papadopoulos, Evangelos P. Markatos, and Nicolar Kourtellis, *No More Chasing Waterfalls: A Measurement Study of the Header Bidding Ad-Ecosystem*, in PROCS. OF THE INTERNET MEASUREMENT CONF. 280 (2019).

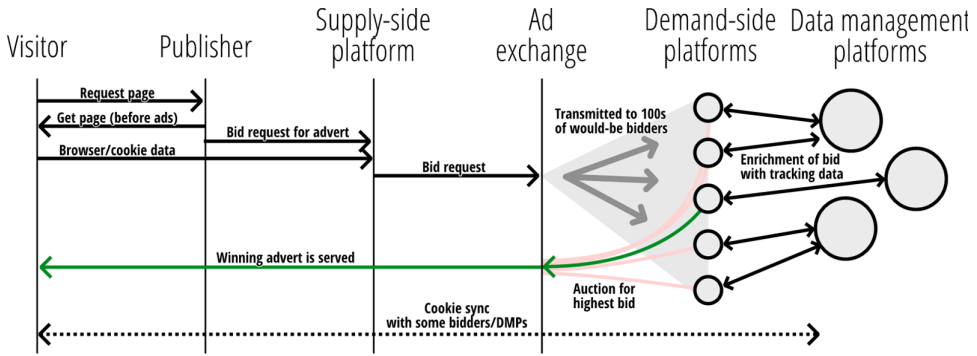

Figure 1. Main actors and processes in RTB (diagram by authors).

A bid request contains a broad array of data about an individual, their device and the website there are visiting. Some of the data in *Authorized Buyers* and *OpenRTB* bid requests relevant to our regulatory discussion include:

- Site
  - URL of the site being visited
  - Site category or topic
- Device
  - Operating system
  - Browser software and version
  - Device manufacturer and model
  - Mobile provider
  - Screen dimensions
- User
  - Unique identifiers set by vendor and/or buyer.
    - AdX's unique person identifier, often from their cookie.
    - The DSP's user identifier, often taken from the cookie of the ADX which has been cookie-synced with a cookie from the DSP's domain.
  - Year of Birth
  - Gender
  - Interests
  - Metadata reporting on consent provided
  - Geography
  - Longitude and latitude
  - Postal/ZIP code

Bid requests with some or all of this information have the potential to directly target individuals in quite granular ways. However, the economic incentives of an auction mean that DSP with more specific knowledge of individuals will win desirable viewers due to being able to target them more specifically and out-bid other entities. As a consequence, the bid request is not the end of the road. The DSP enlists a final actor, the data management platform (DMP). DSPs send bid requests to DMPs, who enrich them by attempting to identify the user in the request and use a variety of data sources, such as those uploaded by the advertiser, collected from other sources, or bought from data brokers. Cambridge Analytica was a notorious DMP, for example, although companies like Google also run DMPs. The DSP with the highest bid not only wins the right to deliver the ad—through the SSP—to the individual. The DSP also wins the right to cookie sync its own cookies with those from the ADX, thus enabling easier linkage of the data to the user's profile in the future.[33]

## C. The GDPR applies to RTB

The GDPR applies to activities that fall within both its material and territorial scope. The GDPR "applies to the processing of personal data wholly or partly by automated means."[34] There are exemptions which self-evidently do not apply in the case of RTB, such as whether the activity if processed "by a natural person in the course of a purely personal or household activity."[35] Therefore the discussion of whether RTB falls within the material scope of the GDPR centers on the GDPR's definition of personal data, relevant case law, and applicable guidance from the European Data Protection Board, where Data Protection Authorities (DPAs) from the twenty-seven EU Member States cooperate. Processing includes almost everything that can be done with personal data,[36] and the definition is so wide that it rarely leads to discussion.

As noted, the main relevant question for material scope is whether RTB involves the use of "personal data." Personal data means

> any information relating to an identified or identifiable natural person (data subject); an identifiable natural person is one who can be identified, directly or indirectly, in particular by reference to an identifier such as a name, an identification number, location data, an online identifier or to one or more factors specific to the physical, physiological, genetic, mental, economic, cultural or social identity of that natural person.[37]

Scholarly work already exists explaining why behavioral advertising constitutes personal data processing under the GDPR and will only be summarized here rather than repeated.[38] Bid requests contain enough data to identify an individual or a device—in practice devices are now primarily individual and not shared—in a number of ways. They generally contain unique identifiers that relate to the ADX, which in turn are connected to different identifiers set by tracking infrastructure run by numerous adtech vendors. Indeed, the winner of a bid can, by design of the protocol, cookie sync and connect their own set of identifiers to the ADX's.[39] It is additionally a common practice to cookie sync with other tracking firms outside of the RTB protocol.[40] Furthermore, the bid request contains such a wide array of personal data beyond explicit identifiers that it is likely to be unique in and of itself, and can serve to fingerprint users.[41] This is even more apparent given the way that industry players collate bid request data.[42]

---

[33] *See generally* MUHAMMAD AHMAD BASHIR, ON THE PRIVACY IMPLICATIONS OF REAL TIME BIDDING 16–17 (2019). Note that the type of sync depends on the AdX; some only provide DSP-specific hashes of the cookies to the DSP, limiting the ability to use this mechanism to link data between DSPs and DMPs, while others provide a common AdX identifier which can make it significantly easier to connect data about the same user. For more, see section B.I.1. Explicit Tracking *supra*.

[34] Regulation (EU) 2016/670, General Data Protection Regulation, art. 2(1), 2016 O.J. (L 119) [hereinafter GDPR].

[35] *Id.* at art. 2(2).

[36] *Id.* at art. 4(2).

[37] *Id.* at art. 4(1) (emphasis added).

[38] *See* Frederik J. Zuiderveen Borgesius, *Singling out People without Knowing Their Names – Behavioural Targeting, Pseudonymous Data, and the New Data Protection Regulation*, 32 COMPUT. L. & SEC. REV. 256 (Apr. 2016).

[39] Panagiotis Papadopoulos, Nicolas Kourtellis, and Evangelos P. Markatos, *Cookie Synchronization: Everything You Always Wanted to Know But Were Afraid to Ask*, (Feb. 2020), https://arxiv.org/abs/1805.10505; Tobias Urban, Denis Tatang, Martin Degeling, Thorsten Holz, and Norbert Pohlmann, *The Unwanted Sharing Economy: An Analysis of Cookie Syncing and User Transparency under GDPR*, (Nov. 2018), https://arxiv.org/abs/1811.08660.

[40] Papadopoulos et al., *supra* note 39; Urban et al., *supra* note 39.

[41] See Section I *supra*.

[42] *See* Rebecca Hill, *French Data Watchdog Withdraws Probe from Location Data Guzzling Adtech Biz Vectaury*, THE REGISTER, Feb. 27, 2019, https://www.theregister.co.uk/2019/02/27/cnil_gdpr_vectaury/.

Even if it requires multiple actors—such as publishers, DSPs, ADXs, and SSPs—to do so, data processing in RTB is designed to identify and profile individual users, and that brings it within the scope of data protection. The Court of Justice of the European Union (CJEU) determined in *Breyer*[43] that the data necessary to identify a user need not all be in the hands of the same actor,[44] and that data would not be personal data only if it met the high barrier that "the identification of the data subject was prohibited by law or practically impossible on account of the fact that it requires a disproportionate effort in terms of time, cost, and man-power, so that the risk of identification appears in reality to be insignificant."[45] In the case of adtech, that seems unlikely to apply, particularly given the ways that industry players rely on contractual controls between hundreds of entities,[46] as well as the prevalence of server-to-server data transfers between players which makes connecting data the norm, rather than the exception.[47]

Furthermore, it is worth adding that some courts have utilized a further test to firmly ground such data as that processed in RTB as personal. The Court of Appeal in England and Wales notes that such "browser generated information" serves to "individuate" the user, in the sense they are singled out, proposing a route to determining whether information is personal data that sits alongside the above linkability analysis by the CJEU in *Breyer*.[48] The CJEU has not needed to consider this type of argument yet, but it is worth noting that "singling out" has entered EU law in the recitals to the GDPR.[49]

The GDPR therefore applies to RTB. It is not the only regime to do so—the ePrivacy Directive has specific rules for tracking technologies, which we will discuss further in section D.I.4. below.[50] As the GDPR applies, we must turn to what it requires of those processing personal data to make such activities lawful.

### D. Legal basis

In this section, we show that European data protection law requires consent of the internet user as a legal basis for RTB practices.[51]

#### I. The GDPR's Requirement for a Legal Basis for Processing

Under the Charter of Fundamental Rights of the European Union, processing personal data is only allowed on the basis of the consent of the data subject or another legal basis laid down by law.[52] The GDPR elaborates, and exhaustively lists six possible legal bases.[53] A data controller—an organization using personal data[54]—may only process personal data on the basis of the data subject's consent, or on one of the other five legal bases. These six legal bases were copied from the 1995 Data Protection Directive with only minor amendments;[55] the requirement for a legal basis has been a key part of EU data protection law for twenty-five years. For the private sector, three legal bases are most relevant: consent, necessity for contractual performance, and the legitimate interests provision. We discuss each of those legal bases in turn, and show that generally, only the data subject's consent can provide a legal basis for personal data processing for RTB.

---

[43]Case C-582/14, Patrick Breyer v. Bundesrepublik Deutschland, ECLI:EU:C:2016:779 (Oct. 19, 2016).

[44]*Id.* at para. 44.

[45]*Id.* at para. 46.

[46]INFO. COMM'R'S OFFICE (ICO), UPDATE REPORT INTO ADTECH AND REAL TIME BIDDING para. 21 (June 2019) https://perma.cc/X7PX-EL3L.

[47]Bashir and Wilson, *supra* note 11.

[48]*See* Vidal-Hall v. Google Inc [2015] EWCA Civ 311, para. 115. The Court further developed this strand of case-law, applying it to transient images in facial recognition systems, in *R (on the Application of Bridges) v South Wales Police* [2020] EWCA Civ 1058, para. 46.

[49]GDPR, *supra* note 34, at recital 26.

[50]Directive 2002/58/EC of the European Parliament and of the Council of 12 July 2002 concerning the processing of personal data and the protection of privacy in the electronic communications sector (Directive on privacy and electronic communications), art 5(3), 2002 O.J. (L 201) [hereinafter ePrivacy Directive].

[51]This section is partly based on, and includes sentences from Frederik J. Zuiderveen Borgesius, *Personal Data Processing for Behavioural Targeting: Which Legal Basis?*, 5 INT'L DATA PRIV. L. 163 (2015).

[52]Charter of Fundamental Rights of the European Union: 2010 O.J. (C83) 389, art. 8(2) [hereinafter CFR].

[53]GDPR, *supra* note 34, at art 6(1). *See also* Case C-13/16, Valsts policijas Rīgas reģiona pārvaldes Kārtības policijas pārvalde v. Rīgas pašvaldības SIA "Rīgas satiksme", ECLI:EU:C:2017:336, para. 25 [hereinafter Rīgas satiksme]; *Breyer*, *supra* note 43, at para. 57 ("Article 7 of Directive 95/46 sets out an exhaustive and restrictive list of cases in which the processing of personal data can be regarded as being lawful.")

### 1. Consent

The GDPR states: "Processing shall be lawful only if and to the extent that at least one of the following applies: ... [T]he data subject has given consent to the processing of his or her personal data for one or more specific purposes."[56] The requirements for valid consent are strict under the GDPR. The GDPR's consent definition says that consent of the data subject means any freely given, specific, informed, and unambiguous indication of the data subject's wishes by which he or she, by a statement or by a clear affirmative action, signifies agreement to the processing of personal data relating to him or her.[57] Article 7, on the conditions for consent, makes the requirements for valid consent even stricter.

The following elements can be deduced from the GDPR's consent definition: Valid consent requires (i) an indication of wishes, which is (ii) specific and informed, and (iii) freely given. We discuss each element in turn.

### 1.1. Indication of Wishes

The most important requirement for valid consent is that a data subject gives an unambiguous indication of their wishes by which they, through a statement or by a clear affirmative action, signifies agreement to the processing of personal data.[58] The GDPR's preamble gives a non-exhaustive list of examples of how a data subject can give an indication of wishes: "A written statement, including by electronic means, or an oral statement."[59] The preamble adds that an indication of wishes "could include ticking a box when visiting an internet website, choosing technical settings for information society services or another statement or conduct which clearly indicates in this context the data subject's acceptance of the proposed processing of his or her personal data."[60]

Under the 1995 Data Protection Directive, data controllers sometimes assumed that a data subject consented if he or she failed to object—an opt-out system.[61] However, an opt-out system could generally not lead to a valid indication of wishes and could thus not lead to valid consent under the Data Protection Directive.[62] The GDPR's consent definition is more explicit than the Directive's, as the GDPR requires "a statement or ... a clear affirmative action" for valid consent.[63] Mere inactivity is not an indication of wishes.

---

[54] For a more precise description, see section D.II.1 *infra*.
[55] *See* Directive 95/46/EC, art. 7, 1995 O.J. (L281) [hereinafter Data Privacy Directive].
[56] GDPR, *supra* note 34, at art. 6(1)(a).
[57] *Id.* at art. 4(11).
[58] *Id.*
[59] *Id.* at recital 32. The "preamble" of EU legislation is a kind of introductory text, consisting of "recitals" which give additional explanations. The Court of Justice of the European Union sometimes refers to recitals in data protection cases. *See, e.g.*, Case C-131/12, Google Spain SL v. Agencia Española de Protección de Datos (AEPD), ECLI:EU:C:2014:317, para. 3 (May 13, 2014) [hereinafter Google Spain]; For more on the role of recitals generally, see Todas Klimas & Jurate Vaiciukaite, *The Law of Recitals In European Community Legislation*, 15 ILSA J. Int'l & Compar. L. 61 (Jan. 2008).
[60] The phrase "information society services" refers, roughly summarized, to internet services. In principle, people can indicate their wishes by choosing technical settings for an internet service. To illustrate: A system like Do Not Track could be developed that enables people to give and withhold consent to online tracking. However, such consent through technical settings must, of course, comply with all the GDPR's requirements for valid consent.
[61] Eleni Kosta, *Construing the Meaning of Opt-Out - An Analysis of the European, U.K. and German Data Protection Legislation*, 1 Eur. Data Prot. L. Rev. 16 (2015).

In a case about cookies, the CJEU confirmed in 2019 that the GDPR "expressly precludes 'silence, pre-ticked boxes, or inactivity' from constituting consent."[64] Other CJEU case law affirms that controllers cannot easily assume consent.[65] To sum up: Opt-out systems cannot be used to obtain valid consent; consent requires a clear expression of will.

### 1.2. Specific and Informed

The GDPR's consent definition also requires that consent be *specific* and *informed*.[66] These two elements are largely overlapping.[67] Article 7 gives additional requirements. It is not acceptable to hide a consent request in the small print of a contract, privacy notice, or other document. The "request for consent shall be presented in a manner which is clearly distinguishable from the other matters."[68] Article 7 also requires that a consent request is presented "in an intelligible and easily accessible form, using clear and plain language."[69] Furthermore, consent must be informed to be valid. A consent request must, at a minimum, disclose the controller's identity, and the processing purpose.[70]

The GDPR's preamble adds about the specificity requirement: "Consent should cover all processing activities carried out for the same purpose or purposes. When the processing has multiple purposes, consent should be given for all of them."[71]

Case law says about the specificity requirement that "consent must be specific, that is to say, connected with a processing operation (or series of processing operations) for precise purposes."[72] In the context of cookies, the CJEU says that specific means that consent "must relate specifically to the processing of the data in question."[73] Moreover, the information provided by the controller must ensure that the "user is in a position to be able to determine easily the consequences of any consent he or she might give and ensure that the consent given is well informed."[74]

### 1.3. Freely Given

Only freely given—and thus voluntary—consent can be valid. Consent is only freely given if the data subject has a genuine choice. Article 7(4) gives guidance regarding this requirement: "When assessing whether consent is freely given, utmost account shall be taken of whether, inter alia, the

---

[62]Id.

[63]GDPR, *supra* note 34, at art. 4(11).

[64]Case C-673/17, Bundesverband der Verbraucherzentralen und Verbraucherverbände—Verbraucherzentrale Bundesverband eV v. Planet49 GmbH, ECLI:EU:C:2019:801, para. 62 (Oct. 1, 2019) [hereinafter Planet49].

[65]Joined Cases C-92/09 and C-93/09, Volker und Markus Schecke and Eifert, ECLI:EU:C:2010:662, para. 62 (Nov. 9, 2010); *see also*, Joined Cases C-92/09 and C-93/09, Volker und Markus Schecke and Eifert, ECLI:EU:C:2010:353, Opinion of AG Sharpston, para. 79 (June 17, 2010).

[66]GDPR, *supra* note 34, at art. 4(11).

[67]*See* ELENI KOSTA, CONSENT IN EUROPEAN DATA PROTECTION LAW 224 (2013) (suggesting that "specific" and "informed" are largely overlapping, and that the requirement of specificity may be superfluous).

[68]GDPR, *supra* note 34, at art. 7(2).

[69]Id. That requirement applies "if the data subject's consent is given in the context of a written declaration which also concerns other matters." Presumably, in other situations a consent request must also be "intelligible," and must also use "clear and plain language." Recital 32 adds that in the online context, a consent request must be straightforward and succinct. The recital says that "[i]f the data subject's consent is to be given following a request by electronic means, the request must be clear [and] concise."

[70]Id. at recital 42.

[71]Id. at recital 32.

[72]*See* Case T-343/13, CN v Parliament, ECLI:EU:T:2015:926, para. 61 (Mar. 5, 2015). This case concerned Regulation 2001/45; that regulation uses a similar consent definition as the GDPR.

[73]Planet49, *supra* note 64, at para. 58.

[74]Id. at para. 74.

performance of a contract, including the provision of a service, is conditional on consent to the processing of personal data that is not necessary for the performance of that contract."[75]

In short, a take-it-or-leave-it choice regarding personal data processing can make consent involuntary and thus invalid. A typical example of such a take-it-or-leave-it choice is a tracking wall, a barrier that visitors can only pass if they consent to tracking by third parties. In the spring of 2020, the European Data Protection Board clarified that tracking walls make consent involuntary and therefore invalid: "In order for consent to be freely given, access to services and functionalities must not be made conditional on the consent of a user to the storing of information, or gaining of access to information already stored, in the terminal equipment of a user (so called cookie walls)."[76]

To summarize: Companies can obtain a legal basis for personal data processing if a data subject gives valid consent. In the following two sections, we show that a data controller cannot rely on other legal bases for RTB.

### 2. Necessity for Contractual Performance

Another legal basis in the GDPR is *necessity* for contract performance. Sometimes a controller can have a legal basis for processing if the processing is necessary for performing a contract.[77] In the words of the GDPR, a data controller can have a legal basis for personal data processing if "processing is necessary for the performance of a contract to which the data subject is party . . . ."[78] For example, a newspaper publisher does not need to obtain consent to process the name and address of a subscriber, as far as these personal data are required to deliver the newspaper to the subscriber's home. The personal data is necessary to deliver the newspaper to the subscriber and thus to fulfil the contract.

Can a contract provide a legal basis for personal data processing for RTB? Almost certainly not. For this provision to apply, the processing must be genuinely necessary for performing the contract. CJEU case law has favored the data subject in interpreting necessity narrowly: "As regards the condition relating to the necessity of processing personal data, it should be borne in mind that derogations and limitations in relation to the protection of personal data must apply only in so far as is strictly necessary."[79]

The necessity requirement is related to proportionality, as confirmed in CJEU case law.[80] The CJEU has said that "the principle of proportionality requires that [measures] be appropriate for

---

[75]The GDPR's preamble makes the requirements for freely given consent even stricter: *see* GDPR, *supra* note 34, at recital 42 ("Consent should not be regarded as freely given if the data subject has no genuine or free choice or is unable to refuse or withdraw consent without detriment."); *See also* GDPR, *supra* note 34, at recital 43.

[76]European Data Protection Board, *Guidelines 05/2020 on Consent under Regulation 2016/679*, at para. 39 (May 4, 2020). Web-wide tracking is not *necessary* for providing a website. *See* Article 29 Data Protection Working Party, *Opinion 03/2013 on Purpose Limitation*, at 46 (Apr. 2, 2013), https://perma.cc/A8S2-3Y94; Article 29 Data Protection Working Party, *Opinion 06/2014 on the Notion of Legitimate Interests of the Data Controller under Article 7 of Directive 95/46/EC*, at 47, (Apr. 9, 2014), https://perma.cc/G35U-7Y8Y.

[77]GDPR, *supra* note 34, at recital 44.

[78]*Id.* at art. 6.

[79]Rīgas Satiksme, *supra* note 53, at para. 30; *See also* Joined Cases C-293/12 and C-594/12, *Digit. Rts. Ir. Ltd. v. Minister for Commc'ns, Marine & Nat. Res.*, ECLI:EU:C:2014:238, para. 50 (Apr. 8, 2014) ("according to the Court's settled case-law . . . derogations and limitations in relation to the protection of personal data must apply only in so far as is *strictly necessary*.") [hereinafter Digital Rights Ireland].

[80]*See* Joined Cases C-203/15 and C-698/15, Tele2 Sverige AB v. Post- och telestyrelsen, ECLI:EU:C:2016:970, para. 96 (Dec. 21, 2016) ("Due regard to the principle of proportionality also derives from the Court's settled case-law to the effect that the protection of the fundamental right to respect for private life at EU level requires that derogations from and limitations on the protection of personal data should apply only in so far as is strictly necessary . . . .") [hereinafter Tele2]; Case C-73/16, Puškár v. Finančné riaditeľstvo Slovenskej republiky, ECLI:EU:C:2017:725, paras. 111–12 (Sept. 27, 2017) ("It is . . . for the referring court to determine whether the establishment of the contested list is necessary for the performance of the tasks carried out in the public interest at issue . . . It is important, in that regard, to ensure that the principle of proportionality is respected. The protection of the

attaining the legitimate objectives pursued ... and do not exceed the limits of what is appropriate and necessary in order to achieve those objectives."[81] In sum, one should not too easily assume that data collection for RTB and targeted advertising is necessary for performing a contract.

Apart from the necessity requirement, there are legal requirements for entering into a contract. It is dubious whether those requirements are met when somebody merely uses a website or an app. From a legal perspective, the main requirement to enter a contract is that both parties want to enter a contract.[82] To illustrate, the Vienna Sales Convention says that "[a] statement made by or other conduct of the offeree indicating assent to an offer is an acceptance. Silence or inactivity does not in itself amount to acceptance."[83] But somebody who visits a website or uses an app rarely intends the wish to enter a contract about tracking or RTB.[84]

Therefore, in most situations, internet users do not enter a contract with companies about trading personal data for ad targeting against the use of a service. Especially if a company collects or uses information about people without them being aware, it is hard to see how those people could have entered a contract with the company.[85] Indeed, the European Data Protection Board says that the legal basis of "necessity for contract performance" is not an appropriate legal basis for data processing for behavioral advertising; consent is always required for such advertising.[86]

### 3. Necessity for the Controller's Legitimate Interests

Another legal basis that a controller can invoke for personal data processing is the *legitimate interests provision*.[87] Roughly summarized, a controller can rely on this provision when personal data processing is necessary for the purposes of the legitimate interests pursued by the controller or by a third party, and those interests outweigh the data subject's interests or fundamental rights. In the words of the GDPR:

> Processing shall be lawful only if and to the extent that at least one of the following applies: ... (f) processing is necessary for the purposes of the legitimate interests pursued by the controller or by a third party, except where such interests are overridden by the interests or fundamental rights and freedoms of the data subject which require protection of personal data, in particular where the data subject is a child.[88]

The CJEU says that the legitimate interests provision implies three cumulative requirements: "[F]irst, the pursuit of a legitimate interest by the data controller or by the third party ... ; second, the need to process personal data for the purposes of the legitimate interests pursued; and third,

---

fundamental right to respect for private life at the European Union level requires that derogations from the protection of personal data and its limitations be carried out within the limits of what is strictly necessary."). For more on proportionality and data protection law see also Lee Andrew Bygrave, Data Privacy Law: An International Perspective, Data Priv. Law 147–50 (2014). *See also* Digital Rights Ireland, *supra* note 79, at para. 69.

[81]Digital Rights Ireland, *supra* note 79, at para. 46; Tele2, *supra* note 80, at paras. 95–107.

[82]Jan Smits, *The Law of Contract*, in Introduction to Law 53, 59 (Jaap Hage and Bram Akkermans eds., 2014).

[83]United Nations Convention on Contracts for the International Sale of Goods (CISG) art. 18(1), U.N., Apr. 11, 1980, 1489 U.N.T.S. 3.

[84]For additional detail, see Zuiderveen Borgesius, *supra* note 51, at 163–76.

[85]*See also* Article 29 Data Protection Working Party, *Letter to Google*, Oct. 16, 2012, www.cnil.fr/fileadmin/documents/en/20121016-letter_google-article_29-FINAL.pdf; Commission Nationale Informatique & Libertés, *Appendix: Google Privacy Policy: Main Findings and Recommendations*, CNIL Oct. 16, 2012, www.cnil.fr/fileadmin/documents/en/GOOGLE_PRIVACY_POLICY-_RECOMMENDATIONS-FINAL-EN.pdf.

[86]*See Opinion 06/2014*, *supra* note 76, at 17. The Working Party adds that "consent should be required, for example, for tracking and profiling for purposes of direct marketing, behavioral advertisement, data-brokering, location-based advertising or tracking-based digital market research." *Id.* at 47.

[87]GDPR, *supra* note 34, at art. 6(1)(f).

[88]*Id.* at art. 6(1).

that the fundamental rights and freedoms of the person concerned by the data protection do not take precedence."[89]

First, are RTB and targeted advertising legitimate interests? Recital 47 gives "direct marketing purposes" as an example of legitimate interests.[90] The recital thus gives an argument in favor of accepting RTB and targeted advertising as legitimate interests. Moreover, RTB companies can invoke their "freedom to conduct a business in accordance with Union law and national laws and practices," as protected by the Charter of Fundamental Rights of the European Union.[91] The Advocate General of the CJEU confirms that online marketing relates to the freedom to conduct a business.[92] The European Data Protection Board emphasizes, logically, that only lawful practices can form a legitimate interest.[93] If RTB brings serious risks for people's privacy and data protection rights, its lawfulness could be questioned. But for now, let us assume that the controller—a company doing RTB—has some legitimate interest.

The second question is: Is the processing necessary to pursue those interests? As noted in the previous section, the necessity hurdle is difficult to overcome. Whether RTB is necessary is debatable. Suppose that the company's interest is making money with online advertising. In that case, there are many other ways of online advertising that do not entail much personal data collection. For example, contextual advertising does not require collecting data about people. Contextual advertising is the practice where ads are adapted to the context, or content, of a web page, exploiting the notion that content preferences may reflect consumer preferences.[94] For instance, ads for local hotels on a website about tourism in Madrid.

However, a company might also argue that it specializes in behavioral advertising or in RTB. A counter argument could be that behavioral advertising is possible without large-scale data collection. Several systems have been developed for, in short, confidential ad targeting.[95] Already ten years ago, researchers developed Adnostic, a browser plug-in that does not involve sharing one's browsing behavior with a company. Adnostic builds a profile based on the user's browsing behavior and uses that profile to target ads—all within the user's device. Minimal information leaves the user's device, as the behavioral targeting happens in the user's browser.[96] Such techniques are entering practice, such as Google's Federated Learning of Cohorts (FLoC) system for microtargeting within Chrome.[97]

Seeing that behavioral advertising is possible without sharing much data with companies, one could argue that large-scale data collection for behavioral advertising is disproportionate and thus not necessary. The requirements for necessity are indeed strict. However, DPAs rarely, if ever, follow that line of reasoning, perhaps because it risks prescribing certain means of processing.[98]

---

[89]Rīgas Satiksme, *supra* note 53, at para. 28.

[90]GDPR, *supra* note 34, at recital 47.

[91]CFR, *supra* note 52, at art. 16.

[92]Case C-131/12, *Google Spain SL v. Agencia Española de Protección de Datos (AEPD)*, ECLI:EU:C:2013:424, para. 95 (June 25, 2013); *See also Opinion 06/2014*, *supra* note 76, at 25 (marketing is a legitimate interest).

[93]*Opinion 06/2014*, *supra* note 76, at 25.

[94]Kaifu Zhang & Zsolt Katona, *Contextual Advertising*, 31 MKTG. SCI. 980 (2012).

[95]Whether these are privacy or data protection, friendly rather than just confidential—in that respect, see Michael Veale, Reuben Binns, and Jef A *When Data Protection by Design and Data Subject Rights Clash*, 8 INT'L DATA PRIV. L. 105 (2018)—or indeed exempt orchestrating organizations from classification as data controllers, as they may remain orchestrating forces following Case C-25/17, Jehovan todistajat, ECLI:EU:C:2018:551, (July 10, 2018), are questions for another time.

[96]Vincent Toubiana, Arvind Narayanan, Dan Boneh, Helen Nissenbaum, and Solon Barocas, *Adnostic: Privacy Preserving Targeted Advertising*, NETWORK & DISTRIBUTED SYS. SYMP. (Mar. 2010), http://crypto.stanford.edu/adnostic/adnostic.pdf.

[97]*See generally* Bennett Cyphers, *Don't Play in Google's Privacy Sandbox*, ELEC. FRONTIER FOUND. (Aug. 30, 2019), https://www.eff.org/deeplinks/2019/08/dont-play-googles-privacy-sandbox-1.

[98]Acquisti, an economist, makes an argument along those lines. Alessandro Acquisti, *The Economics of Personal Data and the Economics of Privacy*, ECON. PERS. DATA & PRIV.: 30 YEARS AFTER THE OECD PRIV. GUIDELINES 42–43 (Dec. 2010), https://www.oecd.org/sti/ieconomy/46968784.pdf. The EDPB indicate a willingness to move in this direction further in recent draft guidance. *See* European Data Protection Board, *Guidelines 8/2020 on the Targeting of Social Media Users*, at para. 47 (Apr. 13, 2021).

Furthermore, current proposals, such as FLoC, require large-scale infrastructural control and coordination, such as shaping a browser and orchestrating a protocol through it, which is not within the power of all data controllers to achieve.[99] Let us then assume, for argument's sake, that some companies engaged in RTB can, in some situations, pass this necessity test.

That would bring us to a third question: Do the data subject's interests outweigh the company's interests? Few, if any, companies engaged in RTB could overcome this hurdle. The data subject's interests include the fundamental rights to privacy and data protection.[100] Case law of the European Court of Human Rights confirms that people have a reasonable expectation of privacy regarding their Internet use.[101] Moreover, surveys consistently show that people see online tracking and related practices as a privacy invasion.[102]

Indeed, the European Data Protection Board suggests that data controllers cannot rely on the legitimate interests provision for personal data processing for targeted advertising: "[C]onsent should be required, for example, for tracking and profiling for purposes of direct marketing, behavioral advertisement, data-brokering, location-based advertising, or tracking-based digital market research."[103] Several authors agree.[104]

The ICO confirmed in a 2019 report that "the nature of the processing within RTB makes it impossible to meet the legitimate interests lawful basis requirements."[105] The DPA adds that "the only lawful basis for 'business as usual' RTB processing of personal data is consent (i.e. processing relating to the placing and reading of the cookie and the onward transfer of the bid request)."[106] Regardless of this array of explicit regulatory guidance on the inappropriateness of this lawful basis, empirical research finds that many RTB vendors in Europe still claim legitimate interest as a lawful ground.[107]

In conclusion, in almost all cases, the data subject's consent is the only available legal basis for personal data processing for RTB and behavioral advertising under data protection law. Even in the far-fetched case that a company can rely on another legal basis for RTB, separate EU law still requires the company to ask consent, namely for the cookies and similar technologies. That cookie consent requirement is the topic for the next section.

---

[99]Note that FLoC is currently under investigation by several competition authorities, including the UK's Competition and Markets Authority and DG COMP.

[100]Joined Cases C-468/10 & C-469/10, Asociación Nacional de Establecimientos Financieros de Crédito (ASNEF) v. Administración del Estado, ECLI:EU:C:2011:777, para. 41 (Nov. 24, 2011); Joined Cases C-465/00, C-138/01, &C-139/01, Rechnungshof v. Österreichischer Rundfunk, ECLI:EU:C:2003:294, ¶ 68 (May 20, 2003); *Google Spain*, *supra* note 59, at para. 74.

[101]Copland v. United Kingdom, 2007 Eur. Ct. H.R. 253, para. 42, https://hudoc.echr.coe.int/eng?i=001-79996.

[102]*See* Joseph Turow, Jennifer King, Chris Jay Hoofnatle, Amy Bleakley, and Michael Hennessy, *Americans Reject Tailored Advertising and Three Activities that Enable It* (Sept. 29, 2009) (Departmental Paper, Annenberg School of Communications), https://perma.cc/Y37K-NTFH. In Europe, seven out of ten people are concerned that companies might use data for new purposes such as targeted advertising without informing them. *See* European Commission, *Special Eurobarometer 359: Attitudes on Data Protection and Electronic Identity in the European Union* (Jun. 2011); Sophie C. Boerman, Sanne Kruikemeier, and Frederik J. Zuiderveen Borgesius, *Online Behavioral Advertising: A Literature Review and Research Agenda*, 46 J. ADVERT. 363 (2017). *See also* Ben Weinshel et al., *Oh, the Places You've Been! User Reactions to Longitudinal Transparency About Third-Party Web Tracking and Inferencing*, 2019 ACM SIGSAC CONF. ON COMPUT. & COMMN'C SEC. 149 (Jun. 2019) (finding that 71.3% of participants considered it "creepy" for "advertising companies to track which websites [they] visit in order to show [them] ads," and 52.9% to 30.6% said the practice was "unfair.").

[103]*Opinion 03/2013*, *supra* note 76, at 46, ("consent should be required, for example, for tracking and profiling for purposes of direct marketing, behavioural advertisement, data-brokering, location-based advertising or tracking-based digital market research.").

[104]*See* Peter Traung, *EU Law on Spyware, Web Bugs, Cookies, Etc., Revisited: Article 5 of the Directive on Privacy and Electronic Communications*, 31 BUS. L. REV. 218 (2010); Lokke Moerel, *Big Data Protection. How to Make the Draft EU Regulation on Data Protection Future Proof* (Feb. 14, 2014). The Dutch government comes to the same conclusion. *See* Authoriteit Persoonsgegevens, *Investigation into the Combining of Personal data by Google: Report of Definitive Findings (Informal English Translation)*, COLLEGE BESCHERMING PERSOONSGEGEVENS, 81 n. 294 (Nov. 2013), https://perma.cc/87U4-Y3CV.

[105]ICO, *supra* note 46, at 18.

[106]*Id.*

[107]Maximilian Hils, Daniel W. Woods, and Rainer Böhme, *Measuring the Emergence of Consent Management on the Web*, PROCS. OF ACM INTERNET MEASUREMENT CONF. 317 (Oct. 2020); Célestin Matte et al., *Purposes in IAB Europe's TCF: Which Legal Basis and How Are They Used by Advertisers?*, in PRIVACY TECHNOLOGIES AND POLICY 163 (Luís Antunes eds. 2020). Both these studies find evidence that the use of legitimate interest as a legal basis is decreasing over time.

### 4. e-Privacy Directive and Consent for Tracking

Apart from the GDPR, the e-Privacy Directive requires consent for the use of tracking cookies and similar technologies. The European e-Privacy Directive says—roughly summarized— that cookies may only be placed after a website visitor has given his or her informed consent, unless those cookies are necessary for communication or to provide a requested service.[108] A website must also ask the visitor consent if third parties—such as advertising networks or social media companies—place cookies on the visitor's computer via the website.[109]

There are two exceptions to this consent rule. First, a website does not need to ask consent if a cookie is placed for the sole purpose of sending communication. For example, if a cookie is needed for the login procedure of an online bank, no consent is required. Second, consent is not required if a cookie is necessary to provide a service requested by the visitor. No consent is therefore required for cookies that are used, for example, for a virtual shopping cart. And no consent is required for a cookie that is placed when a visitor sets his or her language preferences for a website.

For the sake of readability, we speak of cookies, but the e-Privacy Directive applies to many more technologies. The rule applies as soon as a party places information—such as a cookie—on a user's device or reads information from a user's device. The rule therefore also clearly applies to, for example, flash cookies—also called local shared objects—and some forms of device fingerprinting.[110]

Consent in the e-Privacy Directive must be interpreted as consent in the GDPR.[111] Therefore, consent for cookies must comply with the GDPR's strict requirements for consent, for instance, regarding sufficient information.

The CJEU adds that the information provided by the company operating cookies "must be clearly comprehensible and sufficiently detailed so as to enable the user to comprehend the functioning of the cookies employed."[112] Moreover, "the information that the service provider must give to a website user includes the duration of the operation of cookies and whether or not third parties may have access to those cookies."[113] The disclosure "must enable the data subject to be able to determine easily the consequences of any consent he or she might give and ensure that the consent given is well informed."[114]

The CJEU confirms that opt-out systems do not lead to valid consent for cookies: "[C]onsent . . . is not validly constituted if, in the form of cookies, the storage of information or access to information already stored in a website user's terminal equipment is permitted by way of a pre-checked checkbox which the user must deselect to refuse his or her consent."[115]

If somebody gives consent for the placing of a cookie—as required by the e-Privacy Directive, —he or she does not automatically give consent for related personal data processing.[116] So even after a company obtained consent for dropping a cookie on someone's device, the company still needs a legal basis for personal data processing if the company wants to use personal data for RTB or targeted advertising. If a company wants to use a tracking cookie for personal data processing for RTB or targeted advertising, both the privacy's consent requirement and the GDPR's requirement for a legal basis apply. In practice, a company could ask for consent for a cookie and consent for personal data processing in one consent request.[117]

---

[108]ePrivacy Directive, *supra* note 50, at art. 5(3).

[109]Case C-40/17, *Fashion ID GmbH & Co. KG v. Verbraucherzentrale NRW eV.*, 2019 ECLI:EU:C:2019:629, para. 102 (July 29, 2019) [hereinafter Fashion ID].

[110]The rule also applies, for example, when an app reads the contact list on someone's phone. *See* Authoriteit Persoonsgegevens, *Dutch DPA: WhatsApp non-users better protected* (Nov. 3, 2015), https://autoriteitpersoonsgegevens.nl/en/news/dutch-dpa-whatsapp-non-users-better-protected.

[111]e-Privacy Directive, *supra* note 50, at art. 2(f).

[112]Planet49, *supra* note 64, at para. 74.

[113]*Id.*

[114]Case C-61/19, *Orange Romania SA v. ANSPDCP*, ECLI:EU:C:2020:901, para. 40 (Nov. 11, 2020).

[115]Planet49, *supra* note 64, at para. 74.

[116]For more details, see Zuiderveen Borgesius, *supra* note 51; *see also* Planet49, *supra* note 64, at paras. 69–70.

### 5. Lifting the Ban on Using Sensitive Data

In many cases, there is yet another reason why RTB requires the consent of the data subject. As the UK's Information Commissioner's Office (ICO) notes, RTB often concerns the processing of special categories of data, also called sensitive data.[118] Special categories of data are data about, for instance, someone's political opinions, health, or sexual preferences. In principle, the processing of such data is prohibited. The GDPR defines special categories of data as follows: "Processing of personal data revealing racial or ethnic origin, political opinions, religious or philosophical beliefs, or trade union membership, . . . data concerning health or data concerning a natural person's sex life or sexual orientation shall be prohibited."[119]

RTB can lead to the processing of special category data in several situations. For instance, visits to certain websites—like Muslim news or Kosher recipes, for example—can suggest somebody's likely religion. Visits to certain online newspapers can suggest someone's political opinion, and visits to certain sites can give an indication of someone's sexual preferences.

Determining whether special category data is being processed can be a nuanced task under data protection law. At one end of the spectrum, a controller might argue that they never intended to use such categories, nor made any specific variables to capture them, and therefore should be free of the obligations they entail. On the other end, processing of Web browsing data might result in different advertisements for those in groups that the special categories of data represent. Insofar as processing of high-dimensional data such as Web history should be expected to have the potential to reveal latent special category data, there is an argument that it should be treated as such, as were such data to leak, be transferred or misused, it has similar damaging properties to obviously special category data. However, such a view would see special category safeguards—often explicit consent—apply to a huge variety of datasets as, for example, religion may be inferred from names with relatively high accuracy. The challenges in navigating this trade-off are navigated warily by regulators. The ICO in recent guidance appear to indicate that they believe that special category conditions are triggered (i) where the use of a dataset is intended to infer such categories; and (ii) where a proactive assessment, "part of [a controller's] obligation to implement data protection by design and by default," reveals that, in practice, a model "learns to use particular combinations of features that are sufficiently revealing of a special category."[120]

The ban on using special categories of data can only be lifted under certain specific exceptions. For instance, hospitals can process medical data.[121] For adtech and RTB, the only available exception is the data subject's explicit consent.[122] Regulators have struggled to distinguish explicit consent from bog-standard GDPR consent. Implicit consent—as a contrast to active consent—is already prohibited under both the GDPR and the previous Data Protection Directive 1995.[123] So what is left? The European Data Protection Board leans towards written consent, such as a signed or digitally signed document or statement, or a two-stage process with email verification, but do not provide clarity on the matter.[124] At the very least, the consent obligations discussed in this article apply in the case where special category data is, or may, be processed.

---

[117]Article 29 Data Protection Working Party, *Opinion 02/2013 on Apps on Smart Devices*, at 14 (Feb. 27, 2013), https://perma.cc/CDF8-8QA9 ("[t]hough both consent requirements are simultaneously applicable . . . the two types of consent can be merged in practice . . . ").

[118]ICO, *supra* note 46, at 16.

[119]GDPR, *supra* note 34, at art. 9(1).

[120]*Info. Comm'r's Office, Guidance on AI and Data Protection*, ICO, https://ico.org.uk/for-organisations/guide-to-data-protection/key-dp-themes/guidance-on-ai-and-data-protection/.

[121]GDPR, *supra* note 34, at arts. 9(2)(c), 9(2)(i).

[122]*Id.* at art. 9(2)(a).

[123]Planet49, *supra* note 64, at paras. 52–60.

In conclusion, for several reasons, a company doing RTB can only legally do so after the data subject's consent. To avoid misunderstanding, we are not arguing that informed consent is a panacea, nor that consent requirements are the best way to regulate RTB and adtech. Under what conditions less consent-focused privacy and data protection law might protect privacy remains an interesting question—albeit one that falls outside the scope of this article.[125]

## II. Can RTB Comply?

As we argue that the lawful basis for RTB can only be consent, it is relevant how companies might go about obtaining consent. An important feature of consent is that is has to be established in relation to categories of data processed for a particular purpose by a particular controllership arrangement. So far, we loosely used the word 'controller' to refer to organizations using personal data, but we must rectify that sloppiness.

The GDPR is more precise: The controller is the "body which, alone or jointly with others, determines the purposes and means of the processing of personal data."[126] If "two or more controllers jointly determine the purposes and means of processing, they shall be joint controllers."[127]

Lawful bases in general cannot be transmitted between controllers; they must be established by the controller(s) who are undertaking the processing. For example, the CJEU has stated that even where controllers' processing activities are aligned, such as in cases of joint controllerships, each must establish, justify, and pursue their own legitimate interest in order for processing to be lawful.[128] For consent, this is doubly true, as consent is only valid if it is informed. The required information must include "at least ... the identity of the controller and the purposes of the processing for which the personal data are intended."[129] The nature of the consent required has led to a particular institutional innovation in RTB systems—the *consent management platform* (CMPs).

### 1. Consent Management Platforms

This understanding is part of the driver behind a recent trend within web tracking—the emergence of CMPs. Using these code libraries, which are embedded within webpages and, less frequently, within apps, a large number of third parties—the industry prefers the term 'vendors'—simultaneously seek consent from a data subject in one action. Consent management platforms facilitate this single transactional moment, usually through user interface resembling a banner or a barrier. They emerged in early 2018 as the GDPR came into force, with the market characterized by a handful of major players.[130]

The attempt to get simultaneous consent can and does end up with consent sought for hundreds of vendors at once. A recent study used web scraping to look at the five largest CMPs in the field and found a median number of 315 vendors from whom consent is requested at once.[131] At the time of the study in late 2019, the largest CMP by market share, QuantCast, was nearly always configured to request consent for 542 vendors with a single click.[132] The identity of these vendors changes and fluctuates over time.[133]

---

[124]European Data Protection Board, *supra* note 76, at paras. 91–98.
[125]For an argument that online consent is incapable of being rescued, see Elettra Bietti, *Consent as a Free Pass: Platform Power and the Limits of the Informational Turn*, 40 PACE L. REV. 310 (2019).
[126]GDPR, *supra* note 34, at art. 4(7).
[127]*Id.* at art. 26(1).
[128]Fashion ID, *supra* note 109, at para. 96.
[129]GDPR, *supra* note 34, at recital 42.
[130]Hils et al., *supra* note 107, at 324.
[131]Midas Nouwens, Illaria Liccardi, Micahel Veale, David Karger, and Lalana Kagal, *Dark Patterns after the GDPR: Scraping Consent Pop-Ups and Demonstrating Their Influence*, ACM CONF. ON HUMAN FACTORS IN COMPUTING SYS. 5 (Apr. 2020).

The CMP approach has several problems which result in questions around its legality, which we will discuss below.

### 2. Inability to Withdraw Consent as Required by Law

CMPs seem to breach the GDPR's requirement that consent be "as easy to withdraw as to give."[134] Let us have a look at the common standard for CMPs across the industry: The *Transparency and Consent Framework* (the "Framework") coordinated by the IAB. The IAB is the industry body who coordinate the actors in much of the RTB ecosystem, enabling the processing to be carried out through the standardization, monitoring, and continued negotiation of the *OpenRTB* protocol. The IAB also seeks to assure DPAs that RTB is compliant with the law through contractually limiting and shaping the means and purposes of processing for actors within the ecosystem using the Framework. The latest version of the Framework is 2.0, and it is this version that is being analyzed here.

The Framework sets up a system whereby CMPs can be automatically queried by vendors embedded on a website to get the current data protection status of a visitor to that page, such as whether they have been disclosed the identity of that vendor, whether they have expressed their consent to that vendor, and similar variables. The idea is that if such a query to the CMP indicates the vendor is permitted to access and place information on the user's terminal device and process their personal data, the vendor can proceed to do so.

There is a flaw with this system, however. Imagine a user who consents to the processing of personal data of TRACKERA and TRACKERB on WEBSITE1. TRACKERA and TRACKERB query the CMP embedded on the website, which informs them that consent has been established, and as a consequence, cookies are laid and read from their browser by TRACKERA and TRACKERB, and personal data that is collected linked to these identifiers is processed server-side. The user later revisits WEBSITE1, and altering their settings, refuses tracking by TRACKERA and TRACKERB. Again, both trackers query the CMP, which this time tells them they are not permitted to read data from the browser, nor permitted to process personal data on the basis of consent. Neither tracker therefore links the user who withdrew consent to the same, original time that user gave it.

Such a scenario does not pose problems in relation to the ePrivacy Directive, as, in accordance with the law, the trackers did not store or access data on the terminal device following the withdrawal of consent. It does, however, create a problem with the GDPR, as despite the consent being withdrawn, both trackers have not had this result actively communicated to them in relation to that user. They continue to—now illegally—process personal data despite the withdrawal of consent by the user.[135]

Another related problem can be observed when the user moves between websites. Assume the user has consented to TRACKERA processing data on WEBSITE1, but on WEBSITE2, refuses consent to TRACKERA. Under the Framework, WEBSITE2 can store this refusal locally, in its own cookie, rather than updating the global consent string that is stored across websites on a cookie linked to the IAB-managed *consensu.org* server. As a result, the later refusal of consent elsewhere does not pass across websites. Even if it did, it still suffers from the problem described above where the consent refusal in practice only relates to the accessing and reading of data on the terminal device, rather than the server-side processing in the tracking ecosystem.

---

[132]*Id.*

[133]Hu et al., *supra* note 9; Hils et al., *supra* note 107.

[134]GDPR, *supra* note 34, at art. 7(3).

[135]European Data Protection Board, *supra* note 76, para. 117 ("[a]s a general rule, if consent is withdrawn, all data processing operations that were based on consent and took place before the withdrawal of consent—and in accordance with the GDPR—remain lawful, however, the controller must stop the processing actions concerned. If there is no other lawful basis justifying the processing (e.g., further storage) of the data, they should be deleted by the controller.").

The European Data Protection Board notes that "[i]f the withdrawal right does not meet the GDPR requirements, then the consent mechanism of the controller does not comply with the GDPR."[136] Consequently, it seems questionable whether the CMP consent mechanism has provided lawful consent since its introduction by the IAB in 2018 in its first Framework.

### 3. The Impossibility of Global Consent to RTB Infrastructure

The IAB, as part of its coordination function, has put its weight behind a global consent mechanism. Under these proposals, a cookie placed by a CMP via a subdomain of the IAB-controlled *consensu.org* would contain a global consent signal that a CMP accepts as valid across all websites. There is also the recently introduced notion of an "out-of-band" lawful basis, which supposes the possibility of a tracker to obtain a lawful basis outside of the Framework's CMP system.[137]

An analysis of the interaction of consent and joint controllership gives reason to question the legality of this arrangement. The CJEU has held that in the context of embedded Web trackers, key technologies for RTB, a webpage will be a joint controller with the entity processing personal data using this tracker.[138] The predecessor of the European Data Protection Board, the Article 29 Working Party, has said since 2010 that a website publisher and a tracking company are generally joint controllers, if the company operates tracking cookies via that website.[139]

At the time of writing, compliance with the CJEU ruling seems questionable, as companies such as Google still insist they operate independent controllership operations—an interpretation that seems hard to square with the CJEU judgment.[140] Given that the facts specifically concern the technologies and commercial situations of the tracking infrastructures discussed in this article, it seems difficult for companies to question the applicability of the CJEU judgment to online tracking and RTB.

A consequence of the CJEU judgment is that every webpage–tracker combination constitutes a distinct controllership arrangement, even where the tracker company operates across multiple websites in relation to the same data subject. Consent necessary to legitimize this personal data processing—and the interaction with devices under the ePrivacy Directive—relates at least in part to a joint controllership situation. For example, two companies running a research project jointly on the basis of consent could not swap a joint controller out for another, which may not be trusted by the data subject, without re-establishing a lawful basis. If this were possible, it could even be envisaged in stages that a joint controllership arrangement would contain none of the original controllers that established consent in the first instance.

---

[136]Article 29 Working Party, *Guidelines on Consent under Regulation 2016/679*, at 22 (Apr. 10, 2018), http://ec.europa.eu/newsroom/article29/document.cfm?action=display&doc_id=51030.

[137]*See* Interactive Advertising Bureau, *Deprecation of Global Scope Support in TCF*, INTERNET ADVERTISING BUREAU EUR. (June 22, 2021), https://perma.cc/5SRQ-PHKW.

[138]Fashion ID, *supra* note 109, at paras. 84–85.

[139]Article 29 Data Protection Working Party, *Opinion 2/2010 on online behavioural advertising*, at 12 (June 22, 2010) ("publishers will be joint controllers if they collect and transmit personal data regarding their visitors such as name, address, age, location, etc. to the ad network provider.").

[140]*See e.g.* Google, *Tools to Help Publishers Comply with the GDPR*, GOOGLE AD MANAGER HELP, https://support.google.com/admanager/answer/7666366?hl=en. The authors cannot find any discussion of joint controllership in tracker documentation by Facebook, who were implicated directly in Fashion ID. However, Facebook has added a "joint controller addendum" to its Facebook Page product in response to Case C-210/16, Unabhängiges Landeszentrum für Datenschutz Schleswig-Holstein v. Wirtschaftsakademie Schleswig-Holstein GmbH, ECLI:EU:C:2018:388 (Jun. 5, 2018), as have other companies such as LinkedIn (Microsoft) in relation to their similar products. *See Page Insights Controller Addendum*, FACEBOOK, https://www.facebook.com/legal/terms/page_controller_addendum; *LinkedIn Pages Joint Controller Addendum*, LINKEDIN, https://legal.linkedin.com/pages-joint-controller-addendum.

Global consent would hold that a publisher would, instead of collecting consent itself, look to an inherited cookie read by a third-party—the IAB domain *consensu.org*—and assume that consent as applicable to its own joint controllership operation. This, firstly, would directly contradict the finding of the CJEU in *Fashion ID*, which stated that "it is for the operator of the website, rather than for the provider of the social plugin, to obtain that consent, because it is the fact that the visitor consults that website that triggers the processing of the personal data."[141] Furthermore, it would be invalid to equate the consent of one joint controllership operation with another, as in the former, the data subject was informed about a different set of controllers. In this new, separate controllership operation, different controller(s) are in play, and consequently the previous consent is not informed in relation to this operation.

Santos, Bielova, and Matte argue—on the basis of the judgments in *Deutsche Telekom*[142] and *Tele2 and Others*[143]—that consent can be transferred between publishers.[144] They do not elaborate on this argument in detail, but the difference in circumstances and legal context between those judgments and the issue at hand make the argument unconvincing. Both cases relate to a specific aspect of the e-Privacy regime in a telecommunications context where consent is mandatory, relating to whether fresh consent is required to republish the information of a telephone subscriber in a public telephone directory owned by a news organization. The Court leans heavily in *Deutsche Telekom* on the Advocate General's opinion. In teleological analysis, the Advocate General notes that a purpose of the public directory elements of the e-Privacy Directive is to ensure the existence of a comprehensive public directory, and that the provisions on transfer of consent must be interpreted in this context such that this purpose is not "severely compromised."[145] There is no equivalent EU legislation aiming at comprehensive online tracking—indeed, the GDPR specifically highlights online advertising as an example of an application where the "proliferation of actors and the technological complexity of practice" make it hard for data subjects to understand "by whom" data is processed.[146] Taking an opportunity to inform data subjects of this away would appear to go against the specific aims of the law, rather than act in concert with it.

Lastly, it would be unwise for publishers to accept this situation, as they would incur significant liability for invalid consent gathered elsewhere. Global consent IAB cookies can—and are—forged, as the empirical study by Matte, Bielova, and Santos has demonstrated.[147] The publisher accepting global consent would have no proof that consent was ever obtained. The publisher would also be liable as part of a joint controllership operation for a legal action undertaken on the basis of accepting this unverified consent signal, even were it to be theoretically possible to accept as valid. As a recent development, and since this article was initially accepted for publication, the Interactive Advertising Bureau appear to have rolled back their rush towards global consent by deprecating part of the required functionality.[148]

## 4. Too Many Parties for Valid Consent

Modern CMPs operate as to make even "the identity of the controller and the purposes of the processing for which the personal data are intended" too much information to feasibly expect the data subject to be able to read. This is problematic, as consent must be informed—a distinction that is clearer in the French text of the GDPR, which states that consent must be *éclairée* (enlightened or illuminated) rather than simply *informé*, which places the emphasis on the mental state of the data subject rather than the act of having provided information, regardless of its eventual use. Not including the time to operate the interface, or to click on nested privacy policies, a recent empirical study estimated that reading the basic data for all vendors would take on average forty minutes per website.[149] This is clearly not conducive to placing the average data subject in an enlightened position. The Court has stated that information "must enable the data subject to be able to determine easily the consequences of any consent" and "ensure that the consent given is well informed."[150]

---

[141] *Fashion ID*, *supra* note 109, at para. 102.
[142] Case C-543/09, Deutsche Telekom AG v Bundesrepublik Deutschland., ECLI:EU:C:2011:279 (May 5, 2011).
[143] Case C-536/15, Tele2 (Netherlands) BV and Others v Autoriteit Consument en Markt (ACM), ECLI:EU:C:2017:214.
[144] Cristiana Santos, Nataliia Bielova, and Célestin Matte, *Are Cookie Banners Indeed Compliant with the Law?*, Tech. & Regul. 91, 109 (2020).
[145] Case C-543/09, Deutsche Telekom AG v Bundesrepublik Deutschland, ECLI:EU:C:2011:90, at para. 126 (Feb. 17, 2011).
[146] GDPR, *supra* note 34, at recital 58.
[147] Célestin Matte, Nataliia Bielova, and Cristiana Santos, *Do Cookie Banners Respect My Choice? Measuring Legal Compliance of Banners from IAB Europe's Transparency and Consent Framework*, arXiv:1911.09964 (Feb. 2020), https://arxiv.org/abs/1911.09964.
[148] Interactive Advertising Bureau, *supra* note 137.

Moreover, the overarching fairness principle of the GDPR places a focus on creating an enabling environment for autonomous choice and the exercise of data rights, conscious of information asymmetries in the digital environment.[151] Clifford, Graef, and Valcke argue that the fairness principle, and its corresponding manifestation in Article 7, "appears to establish a burden of care on controllers regarding their responsibility to ensure data subjects have been informed and understand the provided information."[152] This is reinforced by the way the EU legislator specified—separately from Article 13 information requirements—the minimal information needed for consent to be informed comprise "at least of the identity of the controller and the purposes of the processing for which the personal data are intended."[153] This seems to be designed assuming that this information, at minimum, must be feasible for the data subject to assess before a consent choice.

As a consequence, with the number of vendors the RTB system requires, CMPs cannot be used to obtain valid consent. This view has recently been echoed by the U.K. Competitions and Markets Authority:

> [I]t is challenging for intermediaries that do not offer user-facing services to obtain consent. At the extreme, this could mean that third-party intermediaries would need to radically reduce the number of other parties they shared a consumer's personal data with to a level the consumer could realistically understand so as to give valid consent to targeted *personalised* advertising.[154]

In conclusion, the GDPR and the ePrivacy Directive require consent for RTB. Current practices by companies engaged in RTB rarely, if ever, lead to valid consent. Indeed, it seems questionable whether it is possible at all to obtain valid consent for RTB.

## E. Transparency
### I. The GDPR's Transparency Requirements

The first of the six overarching principles of EU data protection law is the lawfulness, fairness, and transparency principle. It says that "[p]ersonal data shall be ... processed lawfully, fairly, and in a transparent manner in relation to the data subject."[155] Case law of the CJEU and the European Court of Human Rights confirms the importance of transparency.[156] Since the 1970s,

---

[149]Nouwens et al., *supra* note 131, at 6.
[150]Orange Romania, *supra* note 114, at para. 40.
[151]Damian Clifford & Jef Ausloos, *Data Protection and the Role of Fairness*, 37 Y.B. Eur. L. 130, 139–40 (2018).
[152]Damian Clifford, Inge Graef, and Peggy Valcke, *Pre-Formulated Declarations of Data Subject Consent—Citizen-Consumer Empowerment and the Alignment of Data, Consumer and Competition Law Protections*, 20 German L.J. 679, 685 (2019).
[153]GDPR, *supra* note 34, at recital 42.
[154]Competitions and Markets Authority, *supra* note 31, at 213.
[155]GDPR, *supra* note 34, at art. 5(1)(a).
[156]Tele2, *supra* note 80, at para. 100 ("The fact that the data is retained without the subscriber or registered user being informed is likely to cause the persons concerned to feel that their private lives are the subject of constant surveillance."); Case C-201/14, Smaranda Bara v. Președintele Casei Naționale de Asigurări de Sănătate, ECLI:EU:C:2015:638, para. 33 (Oct. 1, 2015) ("the requirement to inform the data subjects about the processing of their personal data is all the more

transparency has been seen as a core principle for data protection law.[157] It has been suggested that mitigating the abuse of information asymmetry is the data protection law's main goal.[158]

Articles 13 and 14 of the GDPR list information that the data controller must give to the data subject to ensure transparency. The data controller can provide the information, for instance, in a privacy notice on a website.[159] The data controller must give this information regardless of the legal basis for processing.[160] The information that is always required includes: The processing purpose,[161] "the identity and the contact details of the controller,"[162] and "the recipients or categories of recipients of the personal data, if any."[163] Article 12 says that "[t]he controller shall take appropriate measures to provide [such] information to the data subject in a concise, transparent, intelligible, and easily accessible form, using clear and plain language.

The preamble adds that information about personal data processing should be "easily accessible and easy to understand" and "natural persons should be made aware of risks, rules, safeguards and rights in relation to the processing of personal data and how to exercise their rights in relation to such processing."[164] Recital 58 adds that clear information is especially important in the context of online advertising, where the number of actors and the complicated technology may confuse the data subject:

> The principle of transparency requires that any information addressed to the public or to the data subject be concise, easily accessible and easy to understand, and that clear and plain language and, additionally, where appropriate, visualisation be used. Such information could be provided in electronic form, for example, when addressed to the public, through a website. This is of particular relevance in situations where the proliferation of actors and the technological complexity of practice make it difficult for the data subject to know and understand whether, by whom and for what purpose personal data relating to him or her are being collected, such as in the case of online advertising.[165]

---

important since it affects the exercise by the data subjects of their right of access to, and right to rectify, the data being processed."); Bărbulescu v Romania, App. No. 61496/08, para. 133 (Jan. 12, 2016), https://hudoc.echr.coe.int/eng?i=001-159906 ("[t]he Court considers that to qualify as prior notice, the warning from the employer must be given before the monitoring activities are initiated, especially where they also entail accessing the contents of employees' communications. International and European standards point in this direction, requiring the data subject to be informed before any monitoring activities are carried out ....").

[157]See, e.g., Committee of Ministers, *Resolution (73)22 on the protection of the privacy of individuals vis-à-vis electronic data banks in the private sector*, at para. 6 (Sept. 26, 1973) ("[a]s a general rule, the person concerned should have the right to know the information stored about him, the purpose for which it has been recorded, and particulars of each release of this information."). *See also* Committee of Ministers, *Resolution (74)29 on the protection of the privacy of individuals vis-à-vis electronic data banks in the public sector*, at paras. 1, 6 (Sept. 20, 1974) ("as a general rule the public should be kept regularly informed about the establishment, operation and development of electronic data banks in the public sector . . . . [a]s a general rule, the person concerned should have the right to know the information stored about him, the purpose for which it has been recorded, and particulars of each release of this information."). The United States Department of Health, Education, and Welfare mentioned transparency as one of the classic Fair Information Principles in 1973 in U.S. DEP'T HEALTH, EDUC., & WELFARE, RECORDS, COMPUTERS, AND THE RIGHTS OF CITIZENS (1973), https://www.justice.gov/opcl/docs/rec-com-rights.pdf ("there must be no personal-data record-keeping systems whose very existence is secret.").

[158]Paul de Hert and Serge Gutwirth, *Privacy, Data Protection and Law Enforcement: Opacity of the Individual and Transparency of Power*, in PRIVACY AND THE CRIMINAL LAW 61 (2006). *See also* FREDERIK J. ZUIDERVEEN BORGESIUS, IMPROVING PRIVACY PROTECTION IN THE AREA OF BEHAVIOURAL TARGETING § 4.3 (2015).

[159]GDPR, *supra* note 34, at art. 12(1) ("[t]he information shall be provided in writing, or by other means, including, where appropriate, by electronic means.").

[160]Article 13 applies when a firm collects data from the data subject; Article 14 applies where the data have not been obtained from the data subject. Both provisions require largely the same information. The main difference is the moment at which the information must be given.

[161]GDPR, *supra* note 34, at art. 13(1)(c).

[162]*Id.* at art. 13(1)(a).

[163]*Id.* at art. 13(1)(e).

[164]The requirement to inform the data subject about risks is not included in articles 12–14 of the GDPR.

[165]Emphasis added.

The European Data Protection Supervisor has also given guidance regarding the GDPR's transparency requirements.[166]

### II. Can RTB Comply?

Do current RTB practices comply with the GDPR's transparency requirements? And if not, would it be theoretically possible for RTB practices comply with the GDPR's transparency requirements? We suggest that the answers are no in both instances. Currently, all RTB practices seem to be too opaque, and therefore in breach of the GDPR. The ICO notes that "in RTB the privacy information provided often lacks clarity and does not give individuals an appropriate picture of what happens to their data."[167] More worryingly, it seems almost impossible to make RTB comply with the GDPR's transparency requirements.

We start with the clearest transparency requirements of the GDPR. As noted, the controller must always provide "the identity and the contact details of the controller."[168] We recall that website publishers and cooperating RTB companies are joint controllers. If somebody visits a website, the publisher must tell that web user the identity of each joint controller—hence, of each company engaged in RTB concerning that website visit. However, with RTB it is often impossible for the website publisher to predict who will win an auction. Therefore, the publisher does not know in advance which companies—such as advertising networks—will show ads on the site. Neither does the publisher know which companies will collect data via the site.

Indeed, the IAB confirms that it is impossible to tell website visitors in advance which companies will collect his or her data in an RTB scenario. The IAB sent a lobbying document to the European Commission, outlining why the IAB thinks that proposals for a new ePrivacy Regulation—that is supposed to replace the ePrivacy Directive—would mean the end of RTB. The document became public after a freedom of information request.[169] The IAB writes: "

> As it is technically impossible for the user to have prior information about every data controller involved in a real-time bidding (RTB) scenario, programmatic trading, the area of fastest growth in digital advertising spend, would seem, at least prima facie, to be incompatible with consent under GDPR."[170]

Sometimes, website publishers were surprised themselves about which parties were present on their sites. The chairman of the US Association of Online Publishers said, "As a publisher we feel we've been raided by the ad industry. We've done site audits and been flabbergasted by how many third-party cookies have been dropped on our site by commercial partners—they were stealing our data."[171]

In sum, currently, it appears that a website publisher who partners with RTB companies cannot inform visitors about who will collect data about them. Moreover, it appears that it is impossible to inform website visitors about the identity of RTB companies who will collect visitors' data. As noted previously, companies engaged in RTB through a website are generally joint controllers with the website publisher.[172] If the website publisher cannot tell the website visitor the identity of the joint controllers, the publisher cannot comply with the GDPR's requirement to provide "the identity and the contact details of the controller."[173]

---

[166]Article 29 Data Protection Working Party, *supra* note 136.
[167]Information Commissioner's Office, *supra* note 46, at 19.
[168]GDPR, *supra* note 34, at art. 13(1)(a).
[169]Johnny Ryan, *New Evidence to Regulators: IAB Documents Reveal that It Knew that Real-Time Bidding Would Be "Incompatible with Consent under GDPR"*, BRAVE (Feb. 20, 2019), https://perma.cc/7EDM-PJ5S.
[170]IAB Europe, *The EU's Proposed New Cookie Rules: Digital Advertising, European Media, and Consumer Access to Online News, Other Content and Services*, OBTAINED FROM THE EUROPEAN COMMISSION THROUGH FREEDOM OF INFORMATION REQUEST (REF. ARES (2018)6600682), https://perma.cc/PK3H-GYD4.
[171]Hall Emma, *Marketers Could Be Hit by Tough New Data Laws for EU*, ADAGE (Oct. 11, 2013), https://adage.com/article/global-news/marketers-hit-tough-data-laws-eu/244674.
[172]See section D.II.1.

"Given the complexity and opacity of the RTB ecosystem," notes the ICO, "organizations cannot always provide the information required, particularly as they sometimes do not know with whom the data will be shared."[174] The ICO further notes that "RTB also involves the creation and sharing of user profiles within an ecosystem comprising thousands of organizations."[175]

But suppose that one doubts whether all RTB partners must indeed be seen as a joint controller; would such an interpretation make a difference? Probably not. The GDPR requires each controller to tell the data subject "the recipients or categories of recipients of the personal data, if any."[176] If an RTB company cooperates with a website publisher, but should not be seen as a joint controller, the company is a recipient.[177] Following the same logic as above, the publisher cannot tell the data subject who the recipients are, as the publisher does not know in advance who will receive data about the web user. Could a publisher tell the data subject merely the categories of recipients?[178] A publisher might argue that it could comply by saying something like: "We may share your personal data with advertising networks, supply-side platforms, supply-side platforms, and advertising exchanges." It is unlikely that such a statement would comply with the GDPR. As noted, controllers must provide information in a "transparent, intelligible and easily accessible form, using clear and plain language."[179] Further, the GDPR's preamble says that user-friendly information is particularly important "in situations where the proliferation of actors and the technological complexity of practice make it difficult for the data subject to know and understand whether, by whom and for what purpose personal data relating to him or her are being collected, such as in the case of online advertising."[180] Presumably, most website visitors do not know what SSPs are, and do not understand what RTB entails. Indeed, a study on RTB commissioned by the ICO and Ofcom on how users' perceptions of the acceptability of RTB advertising online changed after it was briefly explained to them how it worked. Acceptability initially stood at 63% pre-explanation yet fell to only thirty-six percent after an explanation was provided.[181]

In theory, a publisher could set up a system in which the publisher cooperates with only five companies for displaying ads. In theory, an RTB-like system could be developed, in which those five companies compete in an automated auction. In such a situation, the website publisher could tell the visitor which five companies could collect data about the visitor. Assuming that the publisher can explain for which purposes those companies process the personal data, such a system could perhaps comply with the GDPR's transparency requirements. But for the moment, we are not in this situation.

All in all, it appears that website publishers and companies engaged in RTB generally do not comply with the GDPR's transparency requirements. Moreover, complying with the GDPR's transparency requirements would only be possible if changes were made to RTB practices. Such changes would have to limit, dramatically, the number or parties involved in RTB.

---

[173]GDPR, *supra* note 34, at art. 13(1)(a).

[174]Information Commissioner's Office, *supra* note 46. A Dutch publisher of a website with hundreds of thousands of visitors each month said in an interview, "I don't have any insight into what third parties are collecting on our site. I trust that those companies behave responsibly." Maurits Martijn and Dimitri Tokmetzis, *Big Business is Watching You*, DE CORRESPONDENT, Oct. 8, 2013, https://decorrespondent.nl/66/Big-Business-is-watching-you/3214002-df572412 (our translation).

[175]Information Commissioner's Office, *supra* note 46, at 20.

[176]GDPR, *supra* note 34, at art. 13(1)(e).

[177]*Id.* at art. 4(9) ("recipient means a natural or legal person, public authority, agency or another body, to which the personal data are disclosed, whether a third party or not.").

[178]*Id.* at art. 13(1)(e).

[179]GDPR, *supra* note 34, at art. 12(2).

[180]*Id.* at recital 58.

[181]MICHAEL WORLEDGE AND MIKE BAMFORD, OFF. COMMC'N, ADTECH: MARKET RESEARCH REPORT 19 (Mar. 2019).

## F. Security

### I. The GDPR's Security Requirements

The GDPR's integrity and confidentiality principle could also have been called the security principle. It obliges data controllers to ensure appropriate security for personal data, "including protection against unauthorized or unlawful processing and against accidental loss, destruction or damage, using appropriate technical or organizational measures."[182] Since the early 1970s, data protection laws emphasize the importance of security and confidentiality of data.[183] The CJEU has suggested that security is part of the essence of the fundamental right to the protection of personal data.[184]

The GDPR does not require absolute security; the level of security must be "appropriate."[185] When assessing which level of security is appropriate, controllers and processors should consider "the state of the art, the costs of implementation and the nature, scope, context, and purposes of processing as well as the risk of varying likelihood and severity for the rights and freedoms of natural persons."[186] When assessing the appropriate level of security, the GDPR accounts in particular for "the risks that are presented by processing, in particular from accidental or unlawful destruction, loss, alteration, unauthorized disclosure of, or access to personal data transmitted, stored or otherwise processed."[187] Costs may also be considered when deciding what level of security is appropriate.[188]

CJEU case law also gives some guidance on what should be considered when deciding how much security must be ensured.[189] The CJEU mentions a number of factors to consider when assessing which level of security is appropriate: (i) The quantity of personal data; (ii) the data's sensitivity; and (iii) the risks. The CJEU suggests that a higher level of security is needed "where personal data is subjected to automatic processing;" and "where there is a significant risk of unlawful access to that data."[190]

### II. Can RTB Comply?

Can RTB companies comply with the GDPR's security requirements? For several reasons, the security requirements for RTB are high. We apply the CJEU's elements to assess what level of security is needed.

First, RTB concerns personal data of millions of people. As the ICO notes, "Thousands of organizations are processing billions of bid requests in the UK each week with—at best— inconsistent application of adequate technical and organizational measures to secure the data in transit and at rest."[191]

---

[182]GDPR, *supra* note 34, at art. 5(1)(f). *See also* recitals 39, 78, and 83.

[183]*See, e.g.*, Data Protection Act of the German state of Hesse, Oct. 7, 1970, Gesetz-und Verordnungsblatt für das Land Hessen [Hes GVBl. II] 300-10. Article 2 reads: "The records, data and results covered by data protection shall be obtained, transmitted and stored in such a way that they cannot be consulted, altered, extracted or destroyed by an unauthorized person. This shall be ensured by appropriate staff and technical arrangements." *See also* Article 3. Security was also mentioned in *Resolution (73)22*, *supra* note 157, at paras. 8, 9 and *Resolution (74)29*, *supra* note 157, at paras. 6, 7.

[184]Digital Rights Ireland, *supra* note 79, at para. 40.

[185]GDPR, *supra* note 34, at art. 32(1).

[186]GDPR, *supra* note 34, at art. 32(1).

[187]GDPR, *supra* note 34, at art. 32(2) & recital 83.

[188]*Id.* at art 32. Regarding costs, the CJEU noted in *Digital Rights Ireland* that the Data Retention Directive allowed telecom companies to "to have regard to economic considerations when determining the level of security which they apply, as regards the costs of implementing security measures." The CJEU appeared to disapprove of the possibility to consider costs when assessing the appropriate level of security. Digital Rights Ireland, *supra* note 79, at para. 67.

[189]Digital Rights Ireland, *supra* note 79, at para. 66 ("[The Data Retention] Directive 2006/24 does not lay down rules which are specific and adapted to (i) the vast quantity of data whose retention is required by that directive, (ii) the sensitive nature of that data and (iii) the risk of unlawful access to that data, rules which would serve, in particular, to govern the protection and security of the data in question in a clear and strict manner in order to ensure their full integrity and confidentiality."). The CJEU repeated the three factors in a later judgment. *See, e.g.*, Tele2, *supra* note 80, at para. 122; Joined Cases C-511/18, C-512/18 and C-520/18, La Quadrature du Net v. Premier ministre, ECLI:EU:C:2020:791, para. 132 (Oct. 6, 2020); Case C-623/17, Priv. Int'l v. Sec. State for Foreign & Commonwealth Affs., ECLI:EU:C:2020:790, para. 68 (Oct. 6, 2020).

[190]Case C-362/14, Schrems v Data Prot. Comm'r, ECLI:EU:C:2015:650, para. 91 (Oct. 6, 2015).

Second, the data can be sensitive. For instance, the data can show which websites people visit and when. Case law of the European Court of Human Rights confirms that people have a reasonable expectation of privacy regarding their internet use,[192] and that "information derived from the monitoring of a person's internet use" is covered by the right to private life in article 8 of the European Convention on Human Rights.[193] As noted, someone's website visits may even suggest special categories of data, suggesting one's medical condition, political opinion, or religion, for example.[194]

Moreover, RTB usually involves storing or accessing cookies—or similar files—on the user's device, such as a computer or smart phone. The CJEU says that the right to privacy protects the contents of people's devices: "[A]ny information stored in the terminal equipment of users of electronic communications networks [is] part of the private sphere of the users requiring protection under the European Convention for the Protection of Human Rights and Fundamental Freedoms."[195]

Third, the risks are high. One risk is data leakage. Another risk is that bad actors publish ads, distributed through RTB, to spread malware. Indeed, there are several examples of ads spreading malware, ads that were placed on well-known websites. In sum, the legal security requirements are high in the context of RTB.

Fourth, RTB concerns automated processing, which, according to the CJEU, is a factor that calls for higher security.[196] Fifth, there is a risk of unlawful access to the data. The adtech industry, let alone the data subject, has hardly any control about what happens to people's data during RTB. As the ICO notes in a report on RTB:

> The nature of the processing is what leads to the risk of 'data leakage', which is where data is either unintentionally shared or used in unintended ways. Multiple parties receive information about a user, but only one will 'win' the auction to serve that user an advert. There are no guarantees or technical controls about the processing of personal data by other parties, e.g. retention, security etc. In essence, once data is out of the hands of one party, essentially that party has no way to guarantee that the data will remain subject to appropriate protection and controls.[197]

Many RTB companies could argue, however, that they implement at least one security measure. The GDPR says that controllers—and processors—must implement security measures and gives four examples of possible measures.[198] One of the examples is *pseudonymization*. If a company processes data about individuals but does not know their names, the company can reasonably argue that it only processes pseudonymous data.[199] However, merely pseudonymizing data is not sufficient to comply with the GDPR's security requirements. As the ICO concludes about RTB, "[i]ndividuals have no guarantees about the security of their personal data within the ecosystem."[200] In sum, currently, most RTB practices are breaching three core GDPR requirements, namely the requirements for a legal basis, transparency, and security.

---

[191]Information Commissioner's Office, *supra* note 46, at 23.
[192]Copland, *supra* note 101, at para. 42.
[193]Bărbulescu, *supra* note 156, at para. 72.
[194]GDPR, *supra* note 34, at art. 9; Website visits may suggest one's medical condition (websites about obesity or wheelchair) one's political opinion (certain newspapers), or one's religion (sites with Kosher recipes).
[195]Planet49, *supra* note 64, at para. 70.
[196]Schrems, *supra* note 190, at para. 91.
[197]Information Commissioner's Office, *supra* note 46, at 20–21.
[198]GDPR, *supra* note 34, at recital 78.
[199]*Id*. at art. 4(5) ("pseudonymization means the processing of personal data in such a manner that the personal data can no longer be attributed to a specific data subject without the use of additional information, provided that such additional information is kept separately and is subject to technical and organizational measures to ensure that the personal data are not attributed to an identified or identifiable natural person.").
[200]Information Commissioner's Office, *supra* note 46, at 23.

## G. Discussion

So far, this article focused on positive law: Asking what the law says. We showed that RTB breaches several aspects of the European data protection law. Now we take a step back and explore whether the law makes sense.

In theory, two scenarios are possible. In scenario A, many companies engaged in RTB breach the GDPR. RTB practices are wrong, and the law is right.

In scenario B, RTB illustrates drafting mistakes in the GDPR. In other words, the law is wrong and RTB practices are right. In this scenario, the EU forgot to pay sufficient attention to the adtech industry while drafting the GDPR and adopted rules—the requirements for a legal basis, transparency, and security—that are outdated or otherwise wrong.

In our opinion, we find ourselves in scenario A. Each of the three rules discussed in this article—the requirements for a legal basis, transparency, and security—make sense. For instance, the requirement of a legal basis for processing has been part of EU data protection law for twenty-five years. Apart from the that, the requirement is included in the Charter of Fundamental Rights of the European Union, so there is no serious chance that the requirement is abolished.

RTB companies might try to argue that they should be able to rely on the legitimate interests provision rather than on consent. As explained in section D above, however, we think that argument will not work. In an earlier paper, one of us showed that under the 1995 Data Protection Directive, behavioral advertising can only be based on the legal basis consent.[201] If our claim back then was correct, surely RTB can only be based on consent. RTB is generally riskier and more privacy-invasive than behavioral advertising. Therefore, a claim that RTB can be based on the legitimate interests provision is even more implausible than a claim that behavioral advertising can be grounded on that legal basis.

Is the GDPR's transparency requirement unreasonable? Again, we think not. With good reason, the requirement of transparency regarding personal data usage has been a staple of data protection law since the 1970s. Data protection law aims, among other things, to impede abuse of information asymmetry. Some RTB companies might claim that while transparency in general is a laudable goal, the requirements as specified in the GDPR are too burdensome. We saw that the GDPR requires data controllers to tell their identity to data subjects.[202] In many RTB scenarios, website publishers cannot tell which RTB companies will collect or use data about the website visitors. However, it seems unwise to abolish the requirement that data controllers must disclose their identity. Apart from that, it seems implausible that the EU would abolish that rule in a revision of the GDPR.

Lastly, are the GDPR's security requirements unreasonable? Again, we think not. Data security has been a core tenet of data protection law since the 1970s, and rightly so. It would be ill-advised to abolish or lower the GDPR's security requirements. Apart from that, it is unlikely that the EU would do so. In sum, in our opinion, the non-compliance of RTB with the GDPR is not the fault of the GDPR.

As an aside, RTB companies should not have been surprised that their practices run afoul of the GDPR. The requirements for a legal basis, transparency, and security were included in the 1995 Data Protection Directive too. So, also under the old regime, any data protection lawyer could have told RTB companies that they were on thin ice from a compliance perspective.

### I. Enforcement

How is it possible that such a large breach of the GDPR exists? We briefly highlight a few possible explanations. The compliance deficit in the RTB sector can be largely explained by an enforcement deficit. DPAs have hardly enforced the law in this sector. True, there are exceptions. For instance, the French DPA has given a fine of fifty million euros to Google for not properly explaining what it does with people's personal data.[203] Nevertheless, enforcement against RTB companies is rare. Why is there a lack of enforcement? We suggest a few possibilities.

---

[201] Zuiderveen Borgesius, *supra* note 51, at 163–76.
[202] *See* section E.

First, DPAs are understaffed and overwhelmed. The GDPR applies to uncountable situations in which personal data are used, and DPAs are supposed to oversee compliance in many sectors.

Second, when the 1995 Data Protection Directive still applied—until May 2018—there was more ambiguity about several data protection rules. For instance, RTB companies could— back then—try to argue that the nameless, pseudonymous, data they used fell outside the scope of data protection law. Similarly, RTB companies could have tried to argue that opt-out systems could be used for consent. Such arguments would not have been convincing, but the old rules were vaguer than the GDPR's. Perhaps some DPAs were hesitant to impose sanctions in cases that were likely to lead to long and costly litigation. Several players in the RTB sector have deep pockets and can afford lengthy litigation.

Third, until the GDPR was applicable, many DPAs did not have the power to impose serious fines. Therefore, some companies may have decided that it was rational to make profits in breach of data protection law—after all, the chance of enforcement was low, as was the maximum fine. Alternatively, DPAs may have thought that investigating this complicated sector was not worth the effort.

Fourth, some RTB companies may be established outside Europe, which makes enforcement harder. The GDPR does often apply to companies established outside the EU, but nevertheless, enforcement may be harder than enforcement against a company established in the EU.

Fifth, some companies engaged in RTB are formally established in Ireland. The DPA in Ireland, however, is understaffed and the Irish DPA is sometimes accused of preferring a light touch approach over hard enforcement. Some commentators speak of the "Ireland loophole" in EU data protection law.[204]

Sixth, the GDPR is applicable since May 2018, so perhaps it should not be too much of a surprise if certain sectors are not fully compliant yet.

Seventh, the RTB ecosystem has been left to illegality for so long, it has formed a large, interwoven system that is difficult to regulate using the toolbox of data protection. Data protection legislation draws its heritage from the regulation of databanks, where the controller was clear. In the RTB environment, it is unclear that removing or applying sanctions to any one actor would drive the system into a different state, given the reinforcing effects of the current structure on data collection and sharing practices. Any DPA must act at scale—potentially in relation to many actors at once—which brings daunting issues of capacity, both inside the regulator and in relation to a slow justice system, given the likelihood of appeals in such existential cases for the industry.

While such factors might help to explain the lack of enforcement, the situation is not acceptable. In the long term, alternative business models are needed online to fund journalism, websites, and other services. In the short term, enforcement action must be taken for the sake of the entire legal regime. We note further that some commentators argue that the advertising industry online is unreliable and risks collapse[205]—something which perhaps speaks to the role of data protection principles more broadly, such as accuracy, and the role of data protection in supporting many rights and freedoms online.

DPAs do not have to enforce the law against all companies engaged in RTB. A couple of serious fines against a couple of companies may already help compliance. If one company gets fined, others see that non-compliance can be costly. In many sectors, compliance improved dramatically with data protection law, because the GDPR—unlike the previous law—enabled serious fines. But if DPAs continue non-enforcement, many companies—including those in other sectors—might think that they can break the law with impunity. Some companies may choose to break the law if the expected profit from breaking the law is higher than the chance of being fined multiplied by the expected fine. Therefore, the possibility of high fines is not enough. There must be a credible chance of enforcement.

---

[203]Commission Nationale de l'Informatique et des Libertés, *The CNIL's Restricted Committee Imposes a Financial Penalty of 50 Million Euros against GOOGLE LLC*, CNIL (Jan. 21, 2019), https://www.cnil.fr/en/cnils-restricted-committee-imposes-financial-penalty-50-million-euros-against-google-llc.

[204]Jan-Philipp Albrecht, *#EUDataP: State of the Union*, Speech at the Chaos Comput. Cong. 2013, https://media.ccc.de/v/30C3_-_5601_-_en_-_saal_2_-_201312281400_-_eudatap_state_of_the_union_-_jan_philipp_albrecht.

[205]*See, e.g.*, Tim Hwang, Subprime Attention Crisis: Advertising and the Time Bomb at the Heart of the Internet (2020).

Without enforcement, data protection law risks significant inconsistencies. To take just one, allowing the field of online advertising to get by with significantly problematic practices of consent might have knock-on effects on other sectors who see poor consent practices as legitimate. As a horizontal instrument, enforcement in all areas of the GDPR interact through both regulatory practice and case-law development. Failing to remedy concerns with adtech creates serious risks to the fundamental right of data protection in its entirety. As Lynskey states, "the alternative [is] that data protection becomes part of the problem—a legitimizing framework for exploitative processing practice."[206]

### II. ePrivacy Regulation

The EU has debated additional rules for privacy on the internet, including rules for the adtech sector. In 2017, the European Commission presented a proposal for an ePrivacy Regulation, which should replace the ePrivacy Directive. Especially after amendments by the European Parliament, the proposed ePrivacy Regulation included promising ideas to regulate adtech. For instance, the Parliament suggested to make compliance with Do Not Track and similar signals obligatory for all parties. In such a scenario, people could choose a "do not track me" setting once, in their computer browser, on their phone, or on another device.[207] Through this mechanism, people would not have to give or withhold consent to many different consent requests. But at the moment, it is unclear whether, when, and in what form an ePrivacy Regulation will be adopted—something made more challenging by its close connection to the ongoing and parallel data protection enforcement saga in the area of data retention.[208]

## H. Conclusion

In conclusion, we assessed whether adtech and real-time bidding (RTB) complies with three rules from the GDPR: the requirements for a legal basis, transparency, and security. We showed that for each of the requirements, most RTB practices do not comply. Indeed, it seems close to impossible to make RTB comply.

First, the EU Charter of Fundamental Rights and the GDPR require data controllers, organizations that use personal data, to have a legal basis, such as consent, for that data use. We showed that in virtually all situations, the only available legal basis for RTB is the data subject's prior consent. Moreover, the ePrivacy Directive requires consent for cookies and similar tracking techniques. In practice, RTB companies rarely obtain valid consent. We also showed that it is hard for companies to obtain valid informed consent for RTB. One of the problems is that it is difficult for companies to explain to internet users what will happen to their data in an RTB scenario.

---

[206]Orla Lynskey, *Delivering Data Protection: The Next Chapter*, 21 GERMAN L.J. 80, 84 (2020).
[207]*See generally* Frederik Zuiderveen Borgesius et al., Directorate-General for Internal Policies, Policy Dep't C: Citizen's Rights and Const. Affairs, Eur. Parliament, *An Assessment of the Commission's Proposal on Privacy and Electronic Communications* (May 2017), http://www.europarl.europa.eu/supporting-analyses.
[208]*See generally* Theodore Cristakis & Kenneth Propp, *How Europe's Intelligence Services Aim to Avoid the EU's Highest Court—and What It Means for the United States*, LAWFARE (Mar. 8, 2021), https://www.lawfareblog.com/how-europes-intelligence-services-aim-avoid-eus-highest-court-and-what-it-means-united-states.

Second, the GDPR requires that data controllers are transparent about what will happen to the data subject's data, regardless of they want to obtain the data subject's consent. Controllers must provide clear, plain, and intelligible information. Here, data controllers run into similar problems as under the requirements for informed consent. The data controller must disclose, among other things, its identity to the data subject. If a website publisher cooperates with RTB companies, CJEU case law shows that those companies must be seen as joint controllers. The website publisher must also disclose the identity of all the RTB partners. However, with RTB, a website publisher often does not know in advance who will collect data on its site. The publisher thus cannot disclose the identities of the joint controllers to website visitors. More generally, it seems doubtful whether publishers can ever explain RTB to visitors.

Third, the GDPR requires appropriate security for personal data processing, including protection against unauthorized or unlawful processing. Appropriate security is extra important, as RTB concerns intimate data about millions of people. It seems doubtful whether RTB companies could meet de GDPR's security requirements.

In sum, RTB is difficult to reconcile with core tenets of the GDPR. We call upon DPAs to enforce the GDPR in the adtech sector.

## I. Afterword

During typesetting of this article, in February 2022, the Belgian DPA handed down a decision, supported by the EDPB, concerning the *Transparency and Consent Framework*, the name for the mechanism IAB Europe design and promote which aims at bringing RTB into compliance with the GDPR.[209] In this decision, the DPA explicitly, and on occasion with reference to this article, agrees with many of the points raised here, including around impossibility of relying on legitimate interest; the inability to consent to so many actors; inability to secure data; and the inability to withdraw consent.[210] Several other DPAs, including the Dutch and Danish authorities, have acknowledged the decision in press releases, indicating that further national enforcement around actors in real-time bidding may soon follow.[211] IAB Europe states it intends to appeal the decision.[212]

---

[209]BELGIAN DATA PROTECTION AUTHORITY, *Decision on the Merits 21/2022 of 2 February 2022, Complaint Relating to Transparency & Consent Framework (IAB Europe)*, DOS-2019-01377, https://www.autoriteprotectiondonnees.be/publications/decision-quant-au-fond-n-21-2022-english.pdf (last visited Feb. 2, 2022).

[210]*See generally*, Michael Veale, Midas Nouwens & Cristiana Santos, *Impossible Asks: Can the Transparency and Consent Framework Ever Authorise Real-Time Bidding After the Belgian DPA Decision?*, 2022 TECH. AND REGULATION 12 (2022).

[211]DATATILSYNET, *Belgisk afgørelse kan have betydning for danske hjemmesider*, http://www.datatilsynet.dk/presse-og-nyheder/nyhedsarkiv/2022/feb/belgisk-afgoerelse-kan-have-betydning-for-danske-hjemmesider (last visited Feb. 17, 2022); Jasper Houtman, *Toezichthouder: advertentiebranche moet direct stoppen met online volgen bezoeker*, FD (Feb. 7, 2022), https://fd.nl/tech-en-innovatie/1429434/toezichthouder-advertentiebranche-moet-direct-stoppen-met-online-volgen-bezoeker.

[212]IAB EUROPE, *APD Decision on IAB Europe and TCF*, IAB Europe (Feb. 3, 2022), https://perma.cc/SS32-P6D9.